\documentclass[aps,prb,twocolumn,showpacs,groupedaddress,preprintnumbers,amsmath,amssymb]{revtex4}

\usepackage {amsmath,amssymb,epsfig,color}
\usepackage {graphicx}
\usepackage {dcolumn}
\usepackage {bm}

\begin{document}

\title{A new hybrid LDA and Generalized Tight-Binding method for the electronic structure calculations of strongly correlated electron systems}

\author {V.A.~Gavrichkov}
\author {M.M.~Korshunov}
 \email {mkor@iph.krasn.ru}
\author {S.G.~Ovchinnikov}
 \affiliation {L.V. Kirensky Institute of Physics, Siberian Branch of Russian Academy of Sciences, 660036 Krasnoyarsk, Russia}

\author {I.A. Nekrasov}
\author {Z.V.~Pchelkina}
 \email {pzv@optics.imp.uran.ru}
\author {V.I.~Anisimov}
 \affiliation {Institute of Metal Physics, Russian Academy of Sciences-Ural Division, 620219 Yekaterinburg GSP-170, Russia}

\date{\today}

\begin{abstract}
A novel hybrid scheme is proposed. The {\it ab initio} LDA calculation 
is used to construct the Wannier functions and obtain single electron 
and Coulomb parameters of the multiband Hubbard-type model. In strong 
correlation regime the electronic structure within multiband Hubbard model 
is calculated by the Generalized Tight-Binding (GTB) method, that combines 
the exact diagonalization of the model Hamiltonian for a small cluster (unit cell)
with perturbation treatment of the intercluster hopping and interactions. For undoped 
La$_2$CuO$_4$ and Nd$_2$CuO$_4$ this scheme results in charge transfer insulators with 
correct values of gaps and dispersions of bands in agreement to the ARPES data.
\end{abstract}


\maketitle

\section{Introduction \label{sec_introduction}}

A conventional band theory is based on the density functional theory (DFT) 
\cite{hohenberg1} and on the Local Density Approximation (LDA) \cite{kohn1} within DFT. 
In spite of great success of the LDA for conventional metallic systems it appears to be inadequate 
for strongly correlated electron systems (SCES). 
For instance, LDA predicts La$_2$CuO$_4$ to be a metal whereas, in reality, 
it is an insulator.
Several approaches to include strong correlations in the LDA method are known, 
for example LDA+U \cite{anisimov1} and LDA-SIC \cite{svane1}.
Both methods result in the correct antiferromagnetic insulator ground 
state for La$_2$CuO$_4$ contrary to LDA, but the origin of the 
insulating gap is not correct. It is formed by the local single-electron 
states splitted by spin or orbital polarization. In these approaches the 
paramagnetic phase of the undoped La$_2$CuO$_4$ (above the Neel temperature $T_N$ ) 
will be metallic in spite of strong 
correlation regime $U \gg W$, where $U$ is the Hubbard Coulomb parameter\cite{hubbard1} and $W$ is a 
free electron bandwidth. The spectral weight redistribution between Hubbard 
subbands is very important effect in SCES that is related to the formation 
of the Mott-Hubbard gap in the paramagnetic phase. This effect is 
incorporated in the hybrid LDA+dynamical mean field theory (DMFT) (for review see 
Ref.\cite{anisimov2,Held,psik}) and 
LDA++ approaches. \cite{lichtenstein1} The electron self-energy in LDA+DMFT approach is 
calculated by the DMFT theory in the limit of infinite dimension \cite{voll93,phystoday} and is 
{\bf k}-independent, $\Sigma _{\bf k} \left( E \right) \to \Sigma \left( E \right)$. \cite{metzner1,georges1} 
That is why the correct band dispersion and the ARPES data for High-$T_c$ compounds 
cannot be obtained within LDA+DMFT theory. 
Recent development of the LDA+cluster DMFT method \cite{hettler1, TMrmp} and spectral density functional theory \cite{savrasov1} gives some hopes that non-local corrections may be included in this scheme.

A generalized tight-binding (GTB) \cite{ovchinnikov1} method has been proposed to study the 
electronic structure of SCES as a generalization of Hubbard ideas for 
the realistic multiband Hubbard-like models. The GTB method combines the 
exact diagonalization of the intracell part of the Hamiltonian, construction 
of the Hubbard operators on the basis of the exact intracell multielectron 
eigenstates, and the perturbation treatment of the intercell hoppings and 
interactions. A similar approach to the 3-band $p-d$ model of cuprates \cite{emery1,varma1} 
is known as the cell perturbation method. \cite{lovtsov1,jefferson1,schutler1} The practical 
realization of the GTB method for cuprates required an explicit construction 
of the Wannier functions to overcome the nonorthogonality of the oxygen 
molecular orbitals at the neighboring $CuO_6 $ cells. \cite{gavrichkov1} The GTB 
calculations for undoped and underdoped cuprates are in good agreement with 
ARPES data both in the dispersion of the valence band and in the spectral 
intensity. \cite{gavrichkov1,gavrichkov2} A strong redistribution of spectral weight with hole 
doping and the formation of the in-gap states have been obtained in these 
calculations. Similar GTB calculations for the manganites has been done 
recently. \cite{gavrichkov3}

As any model Hamiltonian approach the GTB method is not {\it ab initio}, there are many 
Hamiltonian parameters like intraatomic energy levels of $p$ and $d$ electrons, 
various $p-d$ and $p-p$ hopping parameters, Coulomb and exchange interaction 
parameters. These parameters have been obtained by fitting the set of 
optical, magnetic \cite{ovchinnikov2} and ARPES \cite{gavrichkov1} data. 
Generally the question arises how unique the set of parameters is.
To overcome this restriction 
we have proposed in this paper a novel LDA+GTB scheme that allows to calculate the 
GTB parameters by the {\it ab initio} LDA approach.

The paper is organized as follows: In Section~\ref{sec_lda_par} 
the construction of Wannier functions from self-consistent LDA eigenfunctions
as well as {\it ab initio} parameters of the multiband $p-d$ model for 
La$_2$CuO$_4$ and Nd$_2$CuO$_4$ are given.
A brief description of 
the GTB method is done in Section~\ref{sec_gtb}. Section~\ref{sec_lda+gtb} contains the 
LDA+GTB band structure calculations for La$_2$CuO$_4$ and Nd$_2$CuO$_4$. 
The effective low-energy $t-J$* 
model with {\it ab initio} parameters is presented in 
Section~\ref{sec_effective_model}. Section~\ref{sec_conclusion} is the conclusion.

\section{Calculation of {\it ab initio} parameters from LDA \label{sec_lda_par}}

To obtain hopping integrals for 
different sets of bands included in consideration we apply 
projection procedure using Wannier functions (WFs) formalism. \cite{wf_anisimov}
WFs were first introduced in 1937 by Wannier \cite{wannier} 
as Fourier transformation of Bloch states $|\psi_{i\bf k}\rangle$
\begin{eqnarray}
\label{WF_psi_def} 
|W_{i}^{\bf T}\rangle& = & \frac{1}{\sqrt{N}}\sum_{\bf k}
e^{-i{\bf kT}}|\psi_{i{\bf k}}\rangle,
\end{eqnarray}
where {\bf T} is lattice translation vector, $N$ is the number of discrete ${\bf k}$ points in the first 
Brillouin zone and $i$ is band index. One major reason why the WFs have seen little practical use 
in solid-state applications is their nonuniqueness
since for a certain set of bands 
any orthogonal linear combination of Bloch functions $|\psi_{i\bf k}\rangle$ 
can be used in (\ref{WF_psi_def}).
Therefore to define them one needs an additional constraint.
Among others Marzari and Vanderbilt \cite{vanderbildt} 
proposed the condition of maximum localization for WFs, 
resulting in a variational procedure. To get a good initial guess
authors of Ref.\onlinecite{vanderbildt} proposed to choose a set of localized 
trial functions $|\phi_n\rangle$ and project them onto the
Bloch states $|\psi_{i\bf k}\rangle$. It was found that this
starting guess is usually quite good. This fact later
led to the simplified calculating scheme 
\cite{pickett} where the variational procedure was abandoned 
as in present work and 
the result of the aforementioned projection was considered as the final 
step.

\subsection{Wannier function formalism \label{wf_form}}
To construct the WFs one should to define a set of trial orbitals $|\phi_n\rangle$ and 
choose the Bloch functions of interest by band indexes (N$_1$, \ldots, N$_2$)
or by energy interval ($E_1, E_2$). 
Non-orthogonalized WFs in reciprocal $|\widetilde{W}_{n\bf k}\rangle$ space
are then the projection of the set of site-centered atomic-like trial orbitals
$|\phi_n\rangle$ 
on the Bloch functions $|\psi_{i\bf k}\rangle$
of the chosen bands: 
\begin{eqnarray}
\label{WF_psi} 
|\widetilde{W}_{n\bf k}\rangle & \equiv & 
\sum_{i(E_1\le \epsilon_{i}({\bf k})\le E_2)} |\psi_{i\bf
k}\rangle\langle\psi_{i\bf k}|\phi_n\rangle,
\end{eqnarray}
where $\epsilon_{i}({\bf k})$ is the band dispersion of $i$-th
band obtained from self-consistent {\it ab initio} LDA calculation. In present work we use
LMT-orbitals \cite {lmto} as trial functions. 
The Bloch functions in LMTO basis are defined as
\begin{eqnarray}
\label{psi} |\psi_{i\bf k}\rangle
& = & \sum_{\mu} c^{\bf k}_{\mu i}|\phi_{\mu}^{\bf k}\rangle,
\end{eqnarray}
where $\mu$ is the combined index representing $qlm$ ($q$ is the atomic number in
the unit cell, $l$ and $m$ are orbital and magnetic quantum numbers), 
$\phi_{\mu}^{\bf k}({\bf r})$ are the Bloch sums of the basis orbitals
$\phi_{\mu}({\bf r-T})$
\begin{eqnarray}
\label{psik} \phi_{\mu}^{\bf k}({\bf r}) & = & \frac{1}{\sqrt{N}}
\sum_{\bf T} e^{i\bf kT} \phi_{\mu}({\bf r}-{\bf T}),
\end{eqnarray}
and the coefficients are
\begin{equation}
\label{coef} c^{\bf k}_{\mu i}=\langle\phi_{\mu}|\psi_{i \bf k}\rangle.
\end{equation}

Since in present work $|\phi_n\rangle$ is an orthogonal LMTO basis set orbital
(in other words $n$ in $|\phi_n\rangle$ corresponds to the particular 
$qlm$ combination), then
$\langle\psi_{i\bf k}|\phi_n\rangle = c_{ni}^{{\bf k}*}$. Hence
\begin{eqnarray}
\label{WF} 
|\widetilde{W}_{n\bf
k}\rangle & = &  \sum_{i=N_1}^{N_2} |\psi_{i\bf k}\rangle
c_{ni}^{{\bf k}*}
 =  \sum_{i=N_1}^{N_2} \sum_{\mu} c^{\bf k}_{\mu i} c_{ni}^{{\bf k}*}
|\phi_{\mu}^{\bf k}\rangle.
\end{eqnarray}
In order to orthonormalize the WFs (\ref{WF}) one needs to calculate the 
overlap matrix $O_{nn'}({\bf k})$
\begin{eqnarray}
\label{O-S} O_{nn'}({\bf k})&\equiv& \langle\widetilde{W}_{n\bf
k}|\widetilde{W}_{n'\bf k}\rangle = \sum_{i=N_1}^{N_2} c^{\bf k}_{ni}
c_{n'i}^{{\bf k}*}, 
\end{eqnarray}
then its inverse square root $S_{nn'}({\bf k})$ is defined as 
\begin{eqnarray}
\label{SS}
S_{nn'}({\bf k}) &\equiv& O^{-1/2}_{nn'}({\bf k}).
\end{eqnarray}
In the derivation of (\ref{O-S}) the orthogonality of Bloch states
$\langle\psi_{n\bf k}|\psi_{n'\bf k} \rangle=\delta_{nn'}$ was used.

From (\ref{WF}) and (\ref{SS}), the orthonormalized WFs in ${\bf k}$-space 
$|W_{n\bf k}\rangle$ can be obtained as
\begin{eqnarray}
\label{WF_orth}\nonumber |W_{n\bf k}\rangle = \sum_{n'} S_{nn'}({\bf k})
|\widetilde{W}_{n'\bf k}\rangle
=\sum_{i=N_1}^{N_2} |\psi_{i\bf k}\rangle\bar{c}_{ni}^{{\bf k}*} &,\\\nonumber
\bar{c}_{ni}^{{\bf k}*}\equiv \langle\psi_{i\bf k}|W_{n\bf k}\rangle
=\sum_{n'} S_{nn'}({\bf k})
c_{n'i}^{{\bf k}*}. 
\end{eqnarray}
Then the matrix element of the  Hamiltonian $\widehat H^{WF}$ in reciprocal
space is
\begin{eqnarray}
\label{E_WF_k} H^{WF}_{nn'}({\bf k}) & = & \langle W_{n\bf
k}|\biggl(\frac{1}{N}\sum_{\bf k'}\sum_{i=N_1}^{N_2} |\psi_{i\bf
k'}\rangle \epsilon_{i}({\bf
k'})\langle\psi_{i\bf k'}|\biggr)|W_{n'\bf k}\rangle \nonumber \\
       & = & \sum_{i=N_1}^{N_2} \bar{c}^{\bf k}_{ni} \bar{c}_{n'i}^{{\bf k}*}
       \epsilon_{i}({\bf k}).
\end{eqnarray}
Hamiltonian matrix element in real space is
\begin{eqnarray}\nonumber
\label{E_WF2} H^{WF}_{nn'}(\bf T) & = & \langle W_{n}^{\bf 0}|
\hat{H} |W_{n'}^{\bf T}\rangle =
\frac{1}{N} \sum_{\bf k}\sum_{i=N_1}^{N_2}
\bar{c}^{\bf k}_{ni} \bar{c}_{n'i}^{{\bf k}*}
\epsilon_{i}({\bf k}) e^{-i\bf kT},
\end{eqnarray}
here atom $n'$ is shifted from its position in the primary unit
cell by a translation vector ${\bf T}$.
For more detailed description of this procedure see. \cite{wf_anisimov}

\subsection{LDA band structure, hopping and Coulomb parameters for p- and n-type cuprates\label{lda_par}}

\begin{figure}
\centering
\includegraphics[clip=true,width=0.35\textwidth,angle=270]{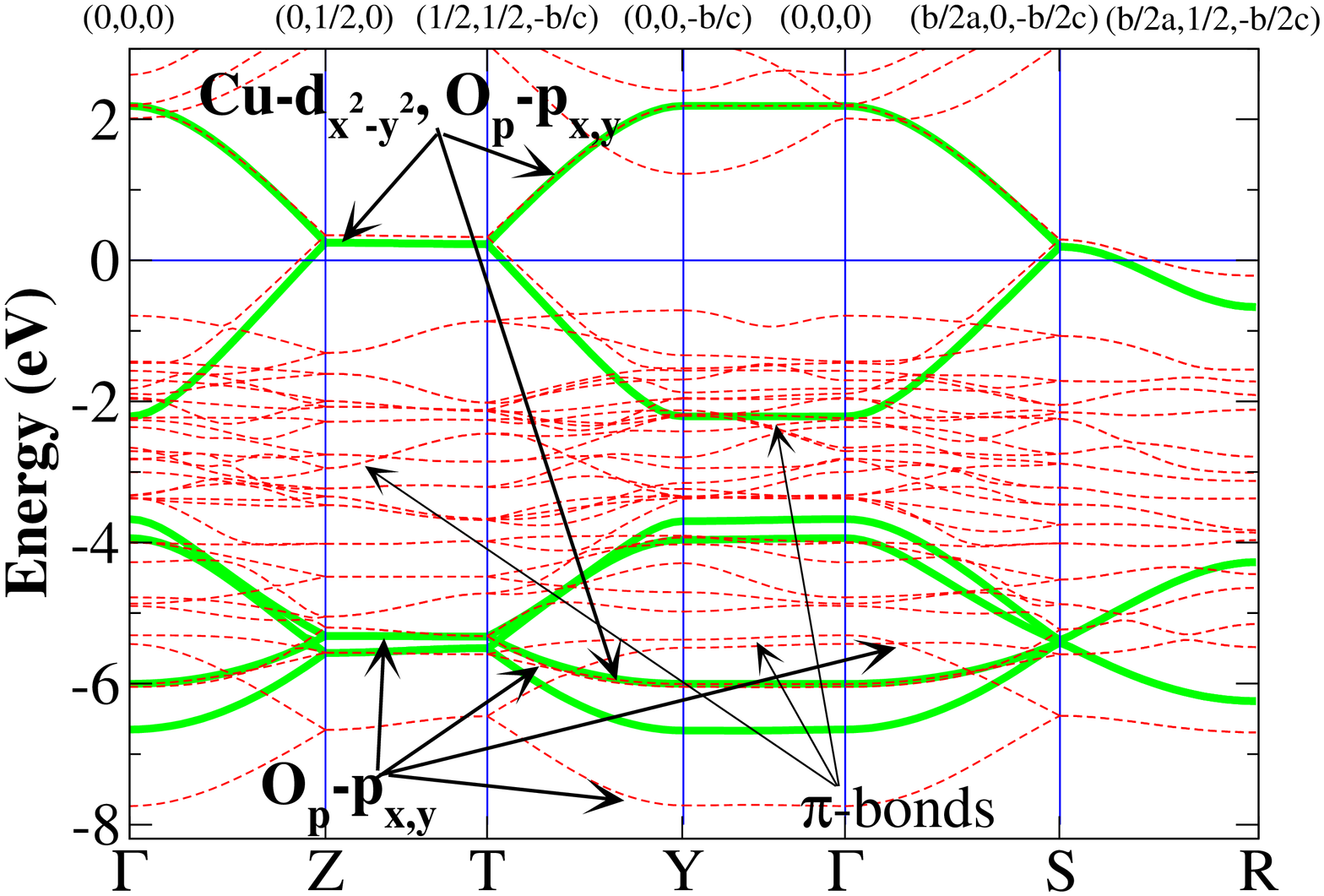}
\caption {\label{la_x2_O} (Color online) Comparison of the band structure of La$_2$CuO$_4$ from LDA 
calculation (dotted lines) and from projection on the Cu-$d_{x^2-y^2}$ and 
O$_p$-$p_x$, O$_p$-$p_y$ set of orbitals (bold solid lines). Fermi level coresponds 
to zero energy.}
\end{figure}
\begin{figure}
\centering
\includegraphics[clip=true,width=0.35\textwidth,angle=270]{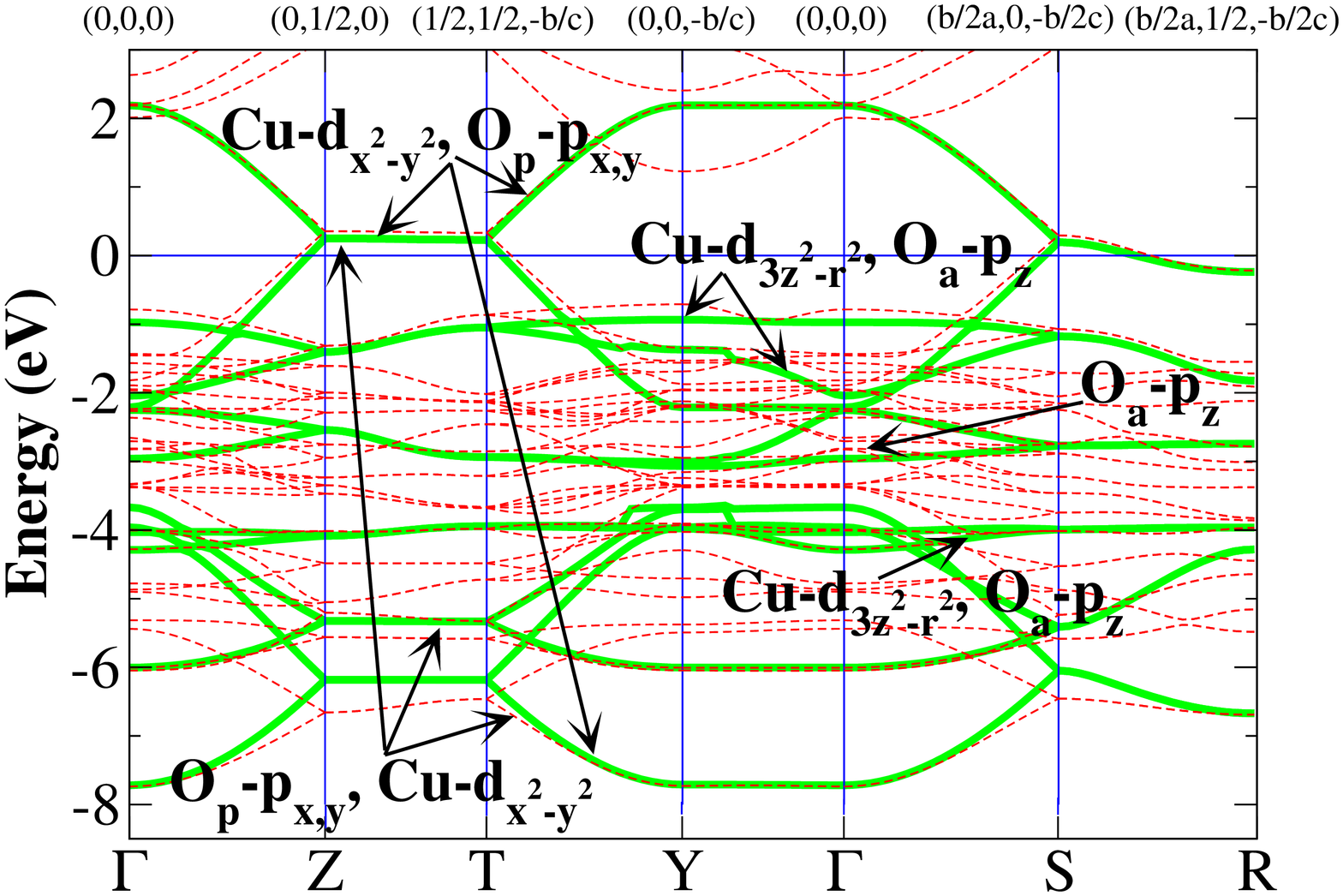}
\caption {\label{la_x2_z2_O} (Color online) The same as in Fig.~\ref{la_x2_O} but  
projection is done on the Cu-$d_{x^2-y^2}$, Cu-$d_{3z^2-r^2}$ and 
O$_p$-$p_x$, O$_p$-$p_y$, O$_a$-$p_z$ set of orbitals.}
\end{figure}
\begin{figure}
\includegraphics[clip=true,width=0.35\textwidth,angle=270]{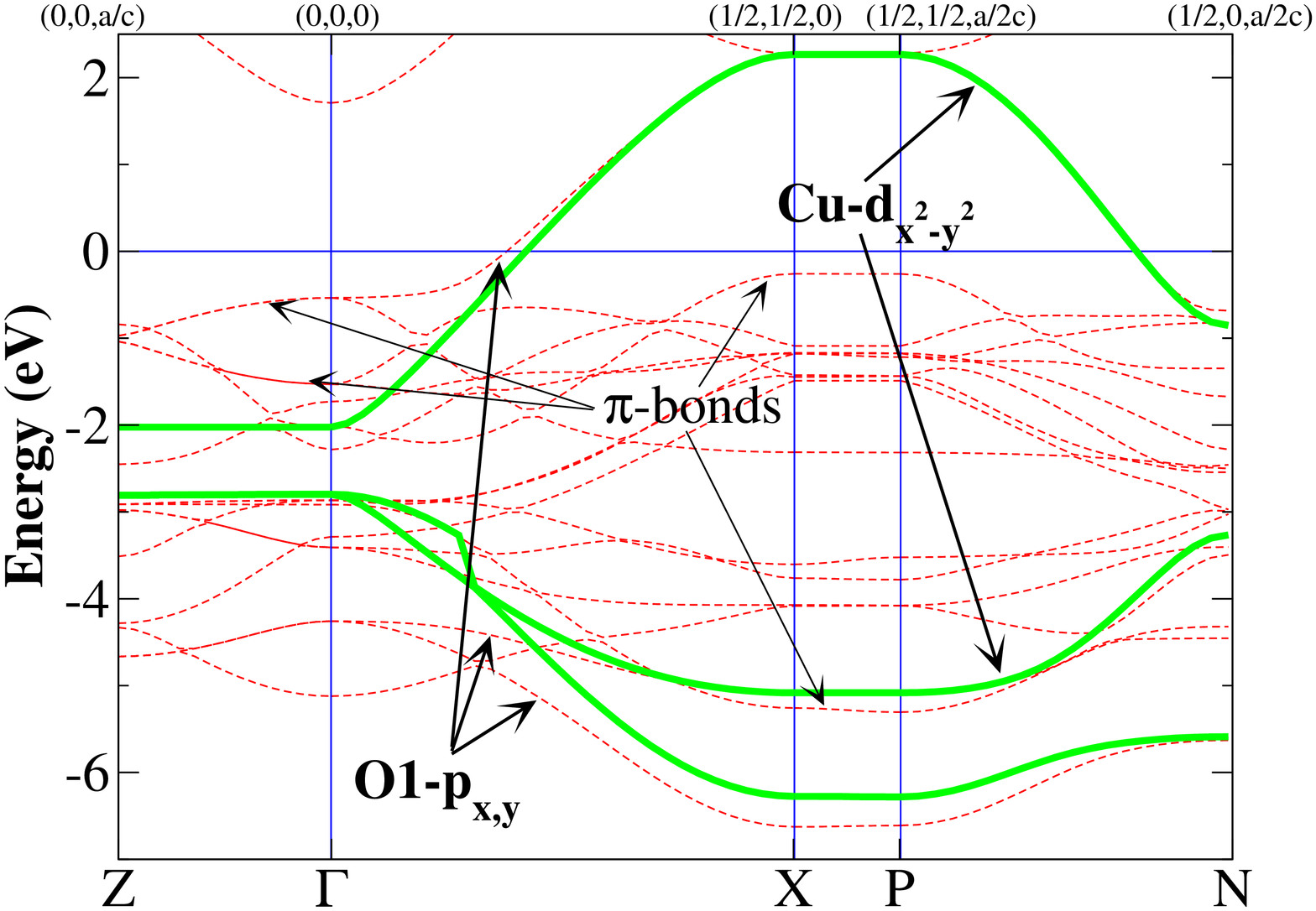}
\caption {\label{nd_x2_O} (Color online) Comparison of the band structure of Nd$_2$CuO$_4$ 
from LDA calculation (dotted lines) 
and from projection on the Cu-$d_{x^2-y^2}$, O1-$p_x$ and 
O1-$p_y$ set of orbitals (bold solid lines). Fermi level corresponds to zero energy.}
\end{figure}
\begin{figure}
\centering
\includegraphics[clip=true,width=0.35\textwidth,angle=270]{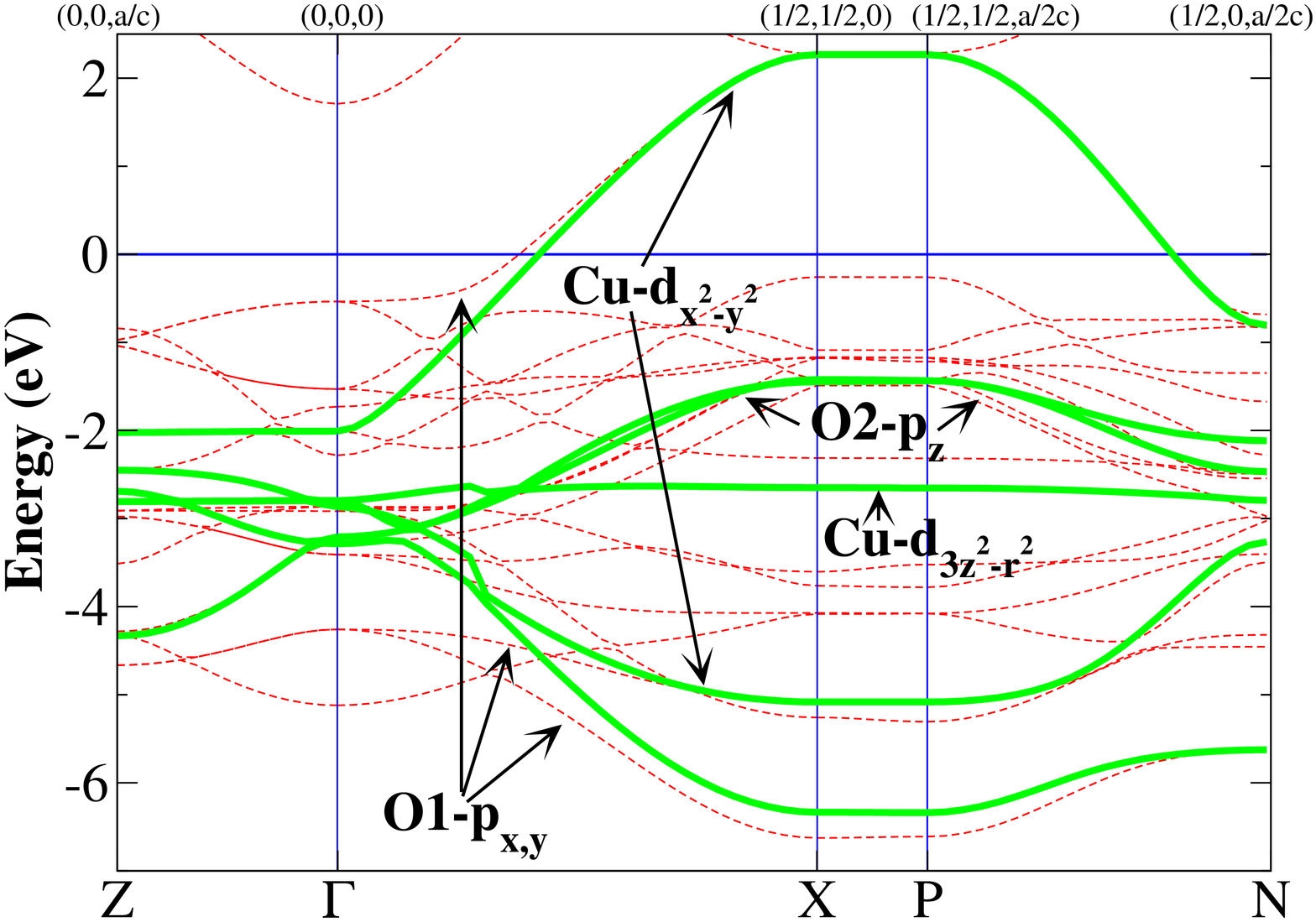}
\caption {\label{nd_x2_z2_O} (Color online) The same as in Fig.~\ref{nd_x2_O} but 
projection is done on the Cu-$d_{x^2-y^2}$, Cu-$d_{3z^2-r^2}$ and 
O1-$p_x$, O1-$p_y$, O2-$p_z$ set of orbitals.}
\end{figure}

Basically all cuprates have one or more CuO$_2$ planes in their structure, 
which are separated by layers of other elements (Ba, Nd, La, ...). They 
provide the carriers in CuO$_2$ plane and according to the type of 
carriers all cuprates can be divided into two classes: p-type and n-type.
In present paper we deal with the simplest representatives of this two classes:
La$_{2-x}$Sr$_x$CuO$_4$ (LSCO) and Nd$_{2-x}$Ce$_x$CuO$_4$ (NCCO) correspondingly.

LDA band calculation for La$_2$CuO$_4$ and Nd$_2$CuO$_4$
was done within LMTO method \cite {lmto} 
using atomic sphere approximation in tight-binding approach \cite{tblmto} (TB-LMTO-ASA). 
In the case of Nd$_2$CuO$_4$ Nd-4$f$ states were treated as pseudocore states.

La$_2$CuO$_4$ at the low temperature and zero doping has the orthorhombic
structure (LTO) with the space group $Bmab$. \cite{strucLa} 
The lattice parameters and atomic cordinates at 10 K were taken from Ref.\onlinecite{strucLa} to be a=5.3346,
b=5.4148 and c=13.1172 \AA, La (0, -0.0083, 0.3616), Cu (0, 0, 0), O$_p$ (0.25, 0.25, -0.0084),
O$_a$ (0, 0.0404, 0.1837). Here and below O$_p$ denotes in-plane oxygen ions and
O$_a$ - apical oxygen ions. In comparison with 
high temerature teragonal structure (HTT) orthorhombic La$_2$CuO$_4$ have
two formula units per unit cell and the CuO$_6$ octahedra are rotated cooperatively 
about the [110] axis. As a result O$_p$ ions are slightly moved off the Cu plane and four
in--plane La-O$_a$ bond lengths are unequal. 

Nd$_2$CuO$_4$ at the room temperature and zero doping has the tetragonal structure 
with the space group $I4/mmm$ \cite{strucNd} 
also called T'-structure. The lattice parameters are a=b=3.94362, 
c=12.1584 \AA. \cite{strucNd} Cu ions at the 2a site (0, 0, 0)
are surrounded by four oxygen ions O1 which occupy 4c position (0, 1/2, 0). The Nd
at the 4e site (0, 0, 0.35112) have eight nearest oxygen ions neighbors O2 at 4d position 
(0, 1/2, 1/4). \cite{strucNd} One can imagine body-centered T'-structure as the HTT structure 
of La$_2$CuO$_4$ but with two oxygen atoms moved from apices of each octahedron to the face of 
the cell at the midpoints between two oxygen atoms on the neighbouring CuO$_2$ planes.
In other words Nd$_2$CuO$_4$ in T'-structure has no apical oxygens around Cu ion.

The LDA band structure of both compounds along the high-symmetry lines in the Brillouin zone
is shown in Figs. \ref{la_x2_O}-\ref{nd_x2_z2_O} by dotted lines. The coordinates of 
high-symmetry points in BZ are given on top of each picture. The complex of bands in the 
energy range (-8, 2.5) eV consists primarily of Cu-$3d$ and O-$2p$ states.
The total bandwidths amount 10 eV for La-cuprate and
7 eV for Nd-cuprate. Contribution of Cu-$3d$ and O-$2p$ orbitals to the different bands 
is displayed by arrows. 

One can see that the band crossing E$_F$ have character of 
Cu-$d_{x^2-y^2}$ and O$_p$-$p_{x,y}$ for La$_2$CuO$_4$ and Cu-$d_{x^2-y^2}$, O1-$p_{x,y}$ 
in the case of Nd$_2$CuO$_4$. It corresponds to antibonding $pd\sigma$ orbital. 
So for hoppings calculation the projection on Cu-$d_{x^2-y^2}$, O$_p$-$p_x$, 
O$_p$-$p_y$ orbitals for La-cuprate and Cu-$d_{x^2-y^2}$, O1-$p_x$, O1-$p_y$ orbitals 
for Nd-cuprate was done.
Such set of orbitals corresponds to the 3-band $p-d$ model. The bands obtained by the 
described in Sec.~\ref{wf_form} projection procedure are shown by solid lines 
in Figs. \ref{la_x2_O} and \ref{nd_x2_O}. It is clearly seen that in case 
of La$_2$CuO$_4$ 3-band model did not reproduce the band crossing E$_F$ properly
(Fig.~\ref{la_x2_O}, SR direction). 

Since 3-band $p-d$ model didn't provide proper description of the LDA bands around Fermi
level the projection on more complex set of trail orbitals for both compounds was done.
The resulting bands are plotted by solid lines in Figs. \ref{la_x2_z2_O} and \ref{nd_x2_z2_O}.
Corresponding multiband $p-d$ model contains Cu-$d_{x^2-y^2}$, Cu-$d_{3z^2-r^2}$, 
O$_p$-$p_x$, O$_p$-$p_y$, O$_a$-$p_z$ states for La$_2$CuO$_4$ and 
Cu-$d_{x^2-y^2}$, Cu-$d_{3z^2-r^2}$, O1-$p_x$, O1-$p_y$, O2-$p_z$ states for Nd$_2$CuO$_4$.
The energy range for projection was (-8.4, 2.5) eV and (-8, 2) eV for the case of  
La-cuprate and Nd-cuprate correspondingly. 
The main effect of taking into account Cu-$d_{3z^2-r^2}$ and O$_a$-$p_z$ 
states for La$_2$CuO$_4$ is the proper description of the band 
structure (in comparison with LDA calculation) at the energies 
up to 2 eV below Fermi level. 
%
From Fig.~\ref{nd_x2_O} and \ref{nd_x2_z2_O} one can see that in case of Nd$_2$CuO$_4$
both sets of trial orbitals properly describe the LDA band crossing 
the Fermi level which has Cu-$d_{x^2-y^2}$ symmetry.
At the same time its bonding part does not agree well with the LDA bands since projection did not
include all Cu-$d$ and O-$p$ orbitals.  

\begin{table}
\caption {Hopping parameters and single electron energies for orthorhombic La$_2$CuO$_4$ obtained 
in WF projection procedure for different sets of
trial orbitals (all values in eV). Here $x^2$, $z^2$, $p_x$, $p_y$, $p_z$ denote Cu-$d_{x^2-y^2}$,
Cu-$d_{3z^2-r^2}$, O$_{p}$-$p_x$, O$_p$-$p_y$, O$_a$-$p_z$ orbitals correspondingly. 
The 3-d and 4-th columns correspond to bases of the 3-band and the multiband $p-d$ models respectively.}
\label{hopsLa}
\begin{ruledtabular}
\begin{tabular}{l|c|c|c}
\footnotesize Hopping             &\footnotesize Connecting            &\footnotesize Cu-$x^2$              &\footnotesize Cu-$x^2$, $z^2$\\
\footnotesize                     &  vector                             &\footnotesize O $p_x$, $p_y$        &\footnotesize O-$p_x$, $p_y$, $p_z$\\
\hline
                                   &                                      &\footnotesize E$_{x^2}$=-1.849      &\footnotesize E$_{x^2}$=-1.849\\                                             
                                   &				          &\footnotesize E$_{p_x}$=-2.767      &\footnotesize E$_{z^2}$=-2.074\\                               
                                   &				          &\footnotesize E$_{p_y}$=-2.767      &\footnotesize E$_{p_x}$=-2.806\\                                     
                                   &				          &                                      &\footnotesize E$_{p_y}$=-2.806\\                            
                                   &				          &                                      &\footnotesize E$_{p_z}$=-1.676\\                             
\hline
\footnotesize t($x^2$,$x^2$)	 &\footnotesize (-0.493,-0.5)          &\footnotesize-0.188                 &\footnotesize-0.188\\
\footnotesize t$'$($x^2$,$x^2$)	 &\footnotesize (-0.985, 0.0)          &\footnotesize 0.001                 &\footnotesize 0.002\\
\hline		
\footnotesize t($z^2$,$z^2$)	 &\footnotesize (-0.493,-0.5)          &	                              &\footnotesize0.054\\
\footnotesize t$'$($z^2$,$z^2$)	 &\footnotesize (-0.985, 0.0)          &	                              &\footnotesize-0.001\\
\hline
\footnotesize t($x^2$,$p_x$)	 &\footnotesize (0.246,0.25,-0.02)     &\footnotesize 1.357                 &\footnotesize 1.355\\
\footnotesize t$'$($x^2$,$p_x$)	 &\footnotesize (-0.739,0.25,-0.02)    &\footnotesize-0.022                 &\footnotesize-0.020\\
\hline
\footnotesize t($z^2$,$p_x$)	 &\footnotesize (0.246,0.25,-0.02)     &	                              &\footnotesize-0.556\\
\footnotesize t$'$($z^2$,$p_x$)	 &\footnotesize (-0.739,0.25,-0.02)    &                                      &\footnotesize-0.028\\
\hline
\footnotesize t($z^2$,$p_z$)	 &\footnotesize (0,0.04,0.445)	       &	                              &\footnotesize0.773\\
\footnotesize t$'$($z^2$,$p_z$)	 &\footnotesize (-0.493,-0.46,-0.445)  &	                              &\footnotesize-0.011\\
\hline
\footnotesize t($p_x$,$p_y$)	 &\footnotesize (0.493, 0.0)	       &\footnotesize-0.841                 &\footnotesize-0.858\\
\footnotesize t$'$($p_x$,$p_y$)  &\footnotesize (0,0.5,0.041)          &\footnotesize 0.775                 &\footnotesize 0.793\\
\footnotesize t$''$($p_x$,$p_y$) &\footnotesize (0.985,0.5,0.041)      &\footnotesize-0.001                 &\footnotesize-0.001\\
\hline
\footnotesize t($p_x$,$p_z$)	 &\footnotesize (-0.246,-0.21,0.465)   &	        	             &\footnotesize-0.391\\
\footnotesize t$'$($p_x$,$p_z$)  &\footnotesize (0.246,0.29,-0.425)    &	        	             &\footnotesize-0.377\\
\footnotesize t$''$($p_x$,$p_z$) &\footnotesize (0.246,-0.21,-0.746)   &	        	             &\footnotesize 0.018\\
\end{tabular}
\end{ruledtabular}
\end{table}

\begin{table}
\caption {Hopping parameters and single electron energies for Nd$_2$CuO$_4$ obtained 
in WF projection procedure for different sets of
trial orbitals (all values in eV). Here $x^2$, $z^2$, $p_x$, $p_y$, $p_z$ denote Cu-$d_{x^2-y^2}$,
Cu-$d_{3z^2-r^2}$, O1-$p_x$, O1-$p_y$, O2-$p_z$ orbitals correspondingly. 
The 3-d and 4-th columns correspond to bases of the 3-band and the multiband $p-d$ models respectively.}
\label{hopsNd}
\begin{ruledtabular}
\begin{tabular}{l|c|c|c}
Hopping            & Connecting              & Cu-$x^2$,      & Cu-$x^2$, $z^2$,\\
                   & vector                  & O $p_x$, $p_y$ & O-$p_x$, $p_y$, $p_z$\\
\hline
                   &                         &\footnotesize E$_{x^2}$=-1.989&\footnotesize E$_{x^2}$=-1.991\\
                   &                         &\footnotesize E$_{p_x}$=-3.409&\footnotesize E$_{z^2}$=-2.778\\
                   &                         &\footnotesize E$_{p_y}$=-3.409&\footnotesize E$_{p_x}$=-3.368\\
		   &                         &                                &\footnotesize E$_{p_z}$=-2.30\\
\hline
\footnotesize t($x^2$,$x^2$)	 &\footnotesize (1, 0)                 &\footnotesize 0.01            &\footnotesize 0.01\\
\footnotesize t$'$($x^2$,$x^2$)	 &\footnotesize (1, 1)                 &\footnotesize -0.00           &\footnotesize-0.00\\
\hline		
\footnotesize t($z^2$,$z^2$)	 &\footnotesize (1, 0)  	       &			      &\footnotesize0.01\\
\footnotesize t$'$($z^2$,$z^2$)	 &\footnotesize (1, 1)  	       &			      &\footnotesize0.00\\
\hline
\footnotesize t($x^2$,$p_x$)	 &\footnotesize  (0.5,0)               &\footnotesize 1.18	      &\footnotesize 1.18\\
\footnotesize t$'$($x^2$,$p_x$)	 &\footnotesize  (0.5,1)               &\footnotesize-0.06	      &\footnotesize-0.06\\
\footnotesize t$'$($x^2$,$p_x$)	 &\footnotesize	 (1.5,0)               &\footnotesize 0.04            &\footnotesize 0.04\\
\footnotesize t$'''$($x^2$,$p_x$)&\footnotesize  (1.5,1)               &\footnotesize 0.00            &\footnotesize 0.00\\
\hline
\footnotesize t($z^2$,$p_x$)	 &\footnotesize  (0.5,0)               &	                      &\footnotesize-0.29\\
\footnotesize t$'$($z^2$,$p_x$)  &\footnotesize  (0.5, 1)              &                              &\footnotesize 0.01\\
\hline
\footnotesize t($z^2$,$p_z$)	 &\footnotesize (0, 0.5, 0.771)	       &	                      &\footnotesize0.10\\
\footnotesize t$'$($z^2$,$p_z$)	 &\footnotesize (1, 0.5, 0.771)	       &	                      &\footnotesize0.02\\
\hline
\footnotesize t($p_x$,$p_y$)	 &\footnotesize (0.5,0.5) 	       &\footnotesize 0.69	      &\footnotesize0.67\\
\footnotesize t$'$($p_x$,$p_y$)  &\footnotesize (1.5,0.5) 	       &\footnotesize 0.00	      &\footnotesize0.00\\
\hline
\footnotesize t($p_x$,$p_z$)	 &\footnotesize (0.5, 0.5, 0.771)      &			      &\footnotesize0.02\\
\footnotesize t$'$($p_x$,$p_z$)  &\footnotesize (0.5,-0.5, 0.771)      &			      &\footnotesize0.02\\
\end{tabular}
\end{ruledtabular}
\end{table}

The resulting hopping parameters and energy of particular orbitals 
for two sets of trial orbitals are presented in Tables ~\ref{hopsLa} and~\ref{hopsNd}. 
The second column contains the connecting vector {\bf T} between two sites.
It is clearly seen that hoppings decay quite rapidly with distance between ions. 

For the multiband $p-d$ model the values of Coulomb parameters are also required.
For Cu in La$_2$CuO$_4$ they were obtained in constrained LDA
supercell calculations \cite{constrain} to be $U=10$~eV and $J=1$~eV. \cite{sorella}
For the Nd$_2$CuO$_4$ we will use the same values of these parameters.

\section{GTB method overview \label{sec_gtb}}

As the starting model that reflects chemical structure of the
cuprates it is convenient to use the 3-band $p-d$ model \cite{emery1,varma1} or the multiband $p-d$ 
model. \cite{gaididei1} While the first one is simplier 
it lacks for some significant features, namely importance of
$d_{z^2}$ orbitals on copper and $p_z$ orbitals on apical oxygen. Non-zero
occupancy of $d_{z^2}$ orbitals pointed out in XAS and EELS experiments
which shows 2-10\% occupancy of $d_{z^2}$ orbitals \cite{bianconi1,romberg1}
and 15\% doping dependent occupancy of $p_z$ orbitals \cite{chen1} in all
hole doped High-$T_c$ compounds). 
Henceforth the multiband $p-d$ model will be used.

Let us consider the Hamiltonian with the following general structure:
\begin {eqnarray}
H &=&\sum\limits_ {f, \lambda, \sigma}(\epsilon_ {\lambda}-\mu) n_ {f
\lambda \sigma} + \sum\limits_ {f \neq g} \sum\limits_ {\lambda, \lambda',
\sigma} T_ {f g}^ {\lambda \lambda'} c_ {f \lambda \sigma}^\dag c_ {g
\lambda' \sigma} \nonumber \\ &+& \frac {1}{2}\sum\limits_ {f, g, \lambda,
\lambda'} \sum\limits_ {\sigma_ {1,2,3,4}} V_ {f g}^{\lambda \lambda'} c_{f \lambda \sigma_1}^\dag c_{f \lambda \sigma_3} c_{g \lambda' \sigma_ 2}^\dag c_{g \lambda' \sigma_4},
\label {eq:Hpd}
\end {eqnarray}
where $c_{f \lambda \sigma}$ is the annihilation operator in Wannier
representation of the  hole at site $f$ at orbital
$\lambda$ with spin $\sigma$,  $n_{f \lambda \sigma}=c_{f \lambda
\sigma}^\dag c_{f \lambda \sigma}$. 

In particular case of cuprates and corresponding multiband $p-d$ model, 
$f$ runs through copper and oxygen sites, index $\lambda$  
run through $d_{x^2-y^2} \equiv d_{x^2}$ and $d_{3z^2-r^2} \equiv d_{z^2}$ orbitals on copper, 
$p_x$ and $p_y$ atomic orbitals on the O$_p$-oxygen sites and $p_z$ orbital
on the apical O$_a$-oxygen; $\epsilon_{\lambda}$ - single-electron energy of the
atomic orbital $\lambda$.  $T_{f g}^{\lambda \lambda'}$ includes matrix
elements of hoppings between copper and  oxygen ($t_{pd}$ for hopping $d_{x^2}
\leftrightarrow p_x,p_y$; $t_{pd}/\sqrt{3}$ for  $d_{z^2} \leftrightarrow
p_x,p_y$; $t'_{pd}$ for $d_{z^2} \leftrightarrow p_z$) and between  oxygen and
oxygen ($t_{pp}$ for hopping $p_x \leftrightarrow p_y$; $t'_{pp}$ for 
hopping $p_x,p_y \leftrightarrow p_z$). The Coulomb matrix elements  $V_{f
g}^{\lambda \lambda'}$ includes intraatomic Hubbard repulsions of two
holes  with opposite spins on one copper and oxygen orbital ($U_d$, $U_p$),
between different  orbitals of copper and oxygen ($V_d$, $V_p$), Hund
exchange on copper and oxygen  ($J_d$, $J_p$) and the nearest-neighbor
copper-oxygen Coulomb repulsion $V_{pd}$.

GTB method \cite{ovchinnikov1,gavrichkov1,gavrichkov2} consist of exact diagonalization of
intracell part of the multiband Hamiltonian (\ref{eq:Hpd}) and perturbative account
of the intercell part.  For La$_{2-x}$Sr$_x$CuO$_4$ and 
Nd$_{2-x}$Ce$_x$CuO$_4$ the unit cells are CuO$_6$ and CuO$_4$ clusters, respectively, 
and a problem of nonorthogonality of the molecular orbitals of adjacent cells 
is solved by an explicit fashion using the diagonalization in {\bf k}-space \cite{raimondi1}.
In a new symmetric basis the intracell part 
of the total Hamiltonian is diagonalized,  allowing to
classify all possible effective quasiparticle excitations in
CuO$_2$-plane  according to a symmetry. 
To describe this process the Hubbard X-operators \cite{hubbard2}
$X_f^m \leftrightarrow X_f^{p,q} \equiv \left| p \right\rangle \left\langle q \right|$ 
are introduced. Index $m \leftrightarrow \left( {p,q} \right)$ enumerates 
quasiparticle with energy 
$\omega _m = \varepsilon _p \left( {N + 1} \right) - \varepsilon _q \left( N \right)$, 
where $\varepsilon_p$ is the $p$-th energy level of the $N$-electron system.
There is a correspondence between Hubbard operators and single-electron creation and annihilation operators:
\begin{equation}
\label{eq:cX}
c_{f\lambda \sigma } = \sum\limits_m \gamma_{\lambda \sigma}\left(m\right) X_{f}^m,
\end{equation}
where $\gamma _{\lambda \sigma } \left( m \right)$ determines the partial weight of a quasiparticle $m$ with spin $\sigma $ and orbital index $\lambda $.
Using this correspondence we rewrite the Hamiltonian (\ref{eq:Hpd})
\begin{equation}
\label{eq:HX}
H=\sum\limits_{f, p} \left(\varepsilon_p-N\mu \right) X^{p,p}_f  + \sum\limits_{f \neq g} \sum\limits_{m,m'} t_{f g}^{m m'} {X_{f}^m}^\dag X_{g}^{m'}.
\end{equation}
This Hamiltonian, actually, have the form of the multiband Hubbard model.

Diagonalization of the Hamiltonian (\ref{eq:Hpd}) mentioned above gives energies $\varepsilon_p$ 
and the basis of Hubbard operators $X_{f}^m$. Values of the hoppings,
\begin{equation}
\label{eq:t_fg}
t_{f g}^{m m'} = \sum\limits_{\sigma, \lambda, \lambda'} T_{f g}^{\lambda \lambda'} \gamma^\ast_{\lambda \sigma}\left(m\right) \gamma_{\lambda' \sigma}\left(m'\right),
\end{equation}
are calculated straightforwardly using the exact diagonalization of the intracell part of the Hamiltonian (\ref{eq:Hpd}).

\begin {figure}
\includegraphics[width=\linewidth]{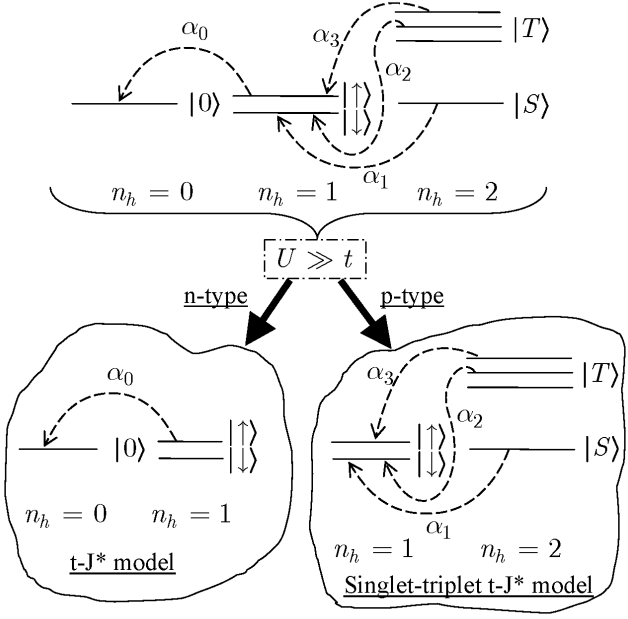}
\caption {\label{fig_model_basis} Schematic picture of states and quasiparticle excitations between them in Hubbard-type model (\ref{eq:HX}). Here $n_h$ stands for number of holes, $\alpha_i$ numerates Fermi-type quasiparticles, states $\left|0\right>$, $\left|\sigma\right>$, $\left|S\right>$, $\left|T\right>$ represents basis of the Hamiltonian (\ref{eq:HX}). Also, bases of effective models are shown. }
\end {figure}

Again, in particular case of multiband $p-d$ model, 
the essential for cuprates multielectron configurations are 
$d^{10}p^6$ (vacuum state $\left| 0 \right\rangle$ in a hole 
representation), single-hole configurations $d^9p^6$, $d^{10}p^5$, and 
two-hole configurations $d^8p^6$, $d^9p^5$, $d^{10}p^4$, $d^{10}p^5p^5$. 
In the single-hole sector of the Hilbert space the $b_{1g}$ molecular orbital, that we will denote later as $\left|\sigma\right> = \left\{ \left|\uparrow\right>, \left|\downarrow\right> \right\}$, has the minimal energy. 
In the two-hole sector the lowest energy states are singlet state $\left|S\right>$ with $^1A_{1g}$ symmetry, that includes Zhang-Rice singlet among other local singlets, and triplet states 
$\left|T\right>=\left|TM\right>$ ($M=+1,0,-1$) with $^3B_{1g}$ 
symmetry. \cite{gavrichkov1,gavrichkov2,raimondi1}
All these states form the basis of the Hamiltonian (\ref{eq:HX}), 
and they are shown together with quasiparticle excitations between them 
in the Fig.~\ref{fig_model_basis}.

In this basis relations (\ref{eq:cX}) between annihilation-creation operators $c_{f\lambda \sigma }$ and 
Hubbard X-operators $X_{f}^m$ are
\begin{eqnarray*}
\label{eq:CvsX}
&&c_{f d_{x^2} \sigma} = u X_f^{0, \sigma}  + 2 \sigma \gamma_x X_f^{\bar \sigma, S}, \\
&&c_{f p_b \sigma} = v X_f^{0, \sigma }  + 2 \sigma \gamma_b X_f^{\bar \sigma, S}, \\
&&c_{f p_a \sigma} = \gamma_a (\sigma \sqrt{2} X_f^{\bar \sigma, T0} - X_f^{\sigma, T2\sigma}), \\
&&c_{f d_{z^2} \sigma} = \gamma_z (\sigma \sqrt{2} X_f^{\bar \sigma, T0} - X_f^{\sigma, T2\sigma}), \\
&&c_{f p_z \sigma} = \gamma_p (\sigma \sqrt{2} X_f^{\bar \sigma, T0} - X_f^{\sigma, T2\sigma}),
\end{eqnarray*}
and the explicit form of the Hamiltonian (\ref{eq:HX}) is given by
\begin{eqnarray}
\label{eq:HXpd}
&&H_{pd} = \sum_{f} \Bigl[ \varepsilon_{1} \sum_{\sigma} X_{f}^{\sigma, \sigma} + \varepsilon_{2S}X_{f}^{S, S} + \varepsilon_{2T} \sum_{M} X_{f}^{TM, TM} \Bigl] \nonumber \\
&&+ \sum_{f \neq g, \sigma} \Bigl[
t_{fg}^{0 0} X_{f}^{\sigma, 0} X_{g}^{0, \sigma} + t_{fg}^{S S} X_{f}^{S, \bar{\sigma}} X_{g}^{\bar{\sigma}, S} \nonumber \\
&&+ 2 \sigma t_{fg}^{0 S} \left( X_{f}^{\sigma, 0} X_{g}^{\bar{\sigma}, S} + h.c. \right) \nonumber \\
&&+ t_{fg}^{S T} \left\{ ( \sigma \sqrt{2} X_{f}^{T0, \bar{\sigma}} - X_{f}^{T2\sigma, \sigma} ) ( v X_{g}^{0, \sigma} + 2\sigma \gamma_{b} X_{g}^{\bar{\sigma}, S} ) + h.c. \right\} \nonumber \\
&&+ t_{fg}^{T T} ( \sigma \sqrt{2} X_{f}^{T0, \bar{\sigma}} - X_{f}^{T2\sigma, \sigma} ) ( \sigma \sqrt{2} X_{g}^{\bar{\sigma}, T0} - X_{g}^{\sigma, T2\sigma} ) \Bigl].
\end{eqnarray}
Here $\bar{\sigma} \equiv -\sigma$. The relation between effective hoppings (\ref{eq:t_fg}) in this Hamiltonian and microscopic parameters of the multiband $p-d$ model is as follows \cite{korshunov2,korshunov3}:
\begin{eqnarray}
\label{eq:t}
t_{fg}^{00} &=& -2t_{pd}\mu_{fg}2uv - 2t_{pp}\nu_{fg}v^2, \nonumber \\
t_{fg}^{SS} &=& -2t_{pd}\mu_{fg}2\gamma_x\gamma_b - 2t_{pp}\nu_{fg}\gamma_b^2, \nonumber \\ t_{fg}^{0S} &=& -2t_{pd}\mu_{fg}(v\gamma_x+u\gamma_b) - 2t_{pp}\nu_{fg}v\gamma_b, \\ t_{fg}^{TT} &=& \frac{2t_{pd}}{\sqrt{3}}\lambda_{fg}2\gamma_a\gamma_z +2t_{pp}\nu_{fg}\gamma_a^2 - 2t'_{pp}\lambda_{fg}2\gamma_p\gamma_a, \nonumber \\ 
t_{fg}^{ST} &=& \frac{2t_{pd}}{\sqrt{3}}\xi_{fg}\gamma_z + 2t_{pp}\chi_{fg}\gamma_a - 2t'_{pp}\xi_{fg}\gamma_p. \nonumber
\end{eqnarray}
The factors $\mu$, $\nu$, $\lambda$, $\xi$, $\chi$ are the coefficients of
Wannier transformation made in the GTB method and $u$, $v$, $\gamma_x$, $\gamma_b$, 
$\gamma_a$, $\gamma_p$, $\gamma_z$ are the matrix elements of annihilation-creation operators in the Hubbard X-operators representation \cite{gavrichkov1}.

Calculations \cite{gavrichkov1,gavrichkov2} of the quasiparticle
dispersion and spectral  intensities in the framework of the multiband $p-d$
model by the GTB method are in very good agreement to the ARPES data on
insulating compound Sr$_2$CuO$_2$Cl$_2$. \cite{wells1,durr1}
Other significant results of this method are \cite{borisov1,borisov2}:

i) pinning of Fermi level in LSCO at low concentrations was
obtained  in agreement with experiments. \cite{ino1,harima1} This pinning
appears due to the in-gap state,  spectral weight of this state is
proportional to doping concentration $x$ and when Fermi level comes to
this in-gap band then Fermi level ``pins'' there. 
The localized in-gap state exist in NCCO also for the same reason as in LSCO, 
but its energy is determined by the extremum of the band at $(\pi/2,\pi/2)$ point 
and it  appears to be above the bottom of the conductivity band. Thus, the first doped 
electron goes into the band state at the $(\pi,0)$ and the chemical potential $\mu$ for
the very small  concentration merges into the band. At higher $x$ it meets
the in-gap state with a pinning at $0.08<x<0.18$ and then $\mu$ again
moves into the band. The dependence $\mu(x)$ for NCCO  is quite
asymmetrical to the LSCO and also agrees with experimental data;
\cite{harima1}

ii) experimentally observed \cite{ino2} evolution of Fermi surface with
doping from hole-type (centered at $(\pi,\pi)$) in the underdoped region to
electron-type (centered at $(0,0)$) in the overdoped region is
qualitatively reproduced;

iii) pseudogap feature for LSCO is obtained as a lowering of
density of states between the in-gap state and the states at the top of the
valence band.

In all these calculations the set of the microscopic model parameters, obtained by fitting to experimental ARPES
data, \cite{wells1,armitage1} was used. Hoppings and single-electron energies are listed in 
Table~\ref{microscopicparams_table}, values of Coulomb parameters are as follows:
\begin{eqnarray} 
\label{eq:Uparams}
\begin{array}{l}
U_d=V_d=9, U_p=V_p=4, \\
J_d=1, J_p=0, V_{pd}=1.5.
\end{array}
\end{eqnarray}

\begin{table}
\caption {\label{microscopicparams_table} Hopping parameters and single-electron 
energies of holes obtained by fitting GTB band structure to experimental data and 
in the {\it ab initio} calculations for p- and n-type cuprates (all values in eV).}
\begin{ruledtabular}
\begin{tabular}{c|cc|cc}
&\multicolumn{2}{c|}{p-type}&\multicolumn{2}{c}{n-type} \\
& fitted & {\it ab initio} & fitted & {\it ab initio} \\
\hline
$\varepsilon_{d_{x^2}}$ & 0    & 0    & 0    & 0    \\
$\varepsilon_{p_x}$     & 1.5  & 0.91 & 1.4  & 1.38 \\
$\varepsilon_{d_{z^2}}$ & 0.2  & 0.14 & 0.5  & 0.79 \\
$\varepsilon_{p_z}$     & 0.45 &-0.26 & 0.45 & 0.31 \\
\hline
$t_{pd}$  & 1    & 1    & 1    & 1    \\
$t_{pp}$  & 0.46 & 0.63 & 0.56 & 0.57 \\
$t'_{pd}$ & 0.58 & 0.57 & 0    & 0.08 \\
$t'_{pp}$ & 0.42 & 0.29 & 0.1  & 0.02 \\
\end{tabular}
\end{ruledtabular}
\end{table}

All results above were obtained treating the intercell hopping in the Hubbard-I approximation. \cite{hubbard1} But the GTB method is not restricted to such a crude approximation. 
The Fourier transform of the two-time retarded Green function in energy representation can be rewritten in terms of matrix Green function $D_{k\sigma }^{m n} \left( E \right) = \left<\left< X_{k\sigma}^m \left| {X_{k\sigma }^n}^\dag \right. \right>\right>_E$:
\begin{equation*}
\left<\left< c_{k\lambda \sigma} \left| c_{k\lambda \sigma}^\dag \right. \right>\right>_E = \sum\limits_{m,m'} \gamma_{\lambda \sigma} \left(m\right) \gamma_{\lambda \sigma}^\ast \left(m'\right) D_{k\sigma }^{m m'} \left( E \right).
\end{equation*}

The diagram technique for Hubbard X-operators is developed \cite{zaitsev1,izumov1} and the generalized Dyson equation \cite{ovchinnikov_book1} reads:
\begin{equation}
\label{eq:D}
\hat {D}_{k\sigma } \left( E \right) = \left\{ {\left[ {\hat {G}_{k\sigma 
}^{(0)} \left( E \right)} \right]^{ - 1} + \hat {\Sigma }_{k\sigma } \left( 
E \right)} \right\}^{ - 1}\hat {P}_{k\sigma } \left( E \right).
\end{equation}

Here, $\hat {\Sigma }_{k\sigma } \left( E \right)$ and $\hat {P}_{k\sigma } \left( E \right)$ are the self-energy and the 
strength operators, respectively. The presence of the strength operator is due to the redistribution of 
the spectral weight, that is an intrinsic feature of SCES. 
First time it was introduced in the spin diagram technique and called ``a strength operator'' \cite{baryakhtar_book1} because the value of $\hat {P}_{k\sigma } \left( E  \right)$ 
determines an oscillator strength of excitations.
It is also should be stressed, that $\hat {\Sigma }_{k\sigma } \left( E \right)$ in Eq.~(\ref{eq:D}) is the self-energy in X-operators representation and therefore it is different from the self-energy entering Dyson equation for the Green function $\left<\left< c_{k\lambda \sigma} \left| c_{k\lambda \sigma}^\dag \right. \right>\right>_E$. The Green function $\hat {G}_{k\sigma }^{(0)} \left( E \right)$ is defined by the formula
\begin{equation}
\label{eq:G0}
\left[ {\hat {G}_{k\sigma }^{(0)} \left( E \right)} \right]^{ - 1} = 
\hat {G}_0^{ - 1} \left( E \right) - 
\hat {P}_{k\sigma } \left( E \right) \hat t_{k\sigma } ,
\end{equation}
where $\hat{G}_0^{ - 1} \left( E \right)$ is the free propagator and $\hat t_{k\sigma } $ is the interaction matrix element 
(for the Hubbard model, $t_{k\sigma }^{m m'} = \gamma_\sigma \left(m\right) \gamma_\sigma^\ast \left(m'\right) t_k$
, and ${G}_0^{m m'} \left( E \right)= \delta_{m m'} / \left( i E - \varepsilon_1 \right)$).

In the Hubbard-I approximation at $U \gg W$ the self-energy $\hat {\Sigma }_{k\sigma } \left( E \right)$ 
is equal to zero and the strength operator $P_{k\sigma }^{m n} \left( E \right)$ is replaced 
by $P_{k\sigma }^{m n} \left( E \right) \to P_{\sigma }^{m n} = \delta _{m n} F_\sigma^m $, 
where $F_\sigma^m = \left\langle {X_f^{p,p} } \right\rangle + \left\langle {X_f^{q,q} } \right\rangle $ is the occupation factor. 
So, in this approximation the following equation is derived from Eq.~(\ref{eq:D}):
\begin{equation}
\label{eq11}
\hat {D}_{k\sigma }^{(0)} = \left\{ {\hat {G}_0^{ - 1} \left( E \right) - 
\hat {P}_{\sigma } \hat t_{k\sigma } } \right\}^{ - 1}\hat {P}_{\sigma }.
\end{equation}

Using diagram technique for the X-operators it is possible to find solution in the GTB method beyond the Hubbard-I approximation. But such discussion is far from the scope of this paper's goals.

It should be stressed that the GTB bands are not free electron bands of the conventional band structure, these are the quasiparticle bands with the number of states in each particular band depending on the occupation number of the initial and final multielectron configurations, and thus on the electron occupation. Bands with zero spectral weight or spectral weight proportional to doping value $x$ appear in the GTB approach.

\section{LDA+GTB method: results and discussion \label{sec_lda+gtb}}

In this Section we will describe the LDA+GTB method itself and some results of this approach.

In LDA+GTB scheme all parameters of the multiband model are calculated  within the 
{\it ab initio} LDA (by Wannier function projection technique, see Sec. \ref{wf_form}) and 
constrained LDA method. \cite{constrain} 
Analysis  of  the  LDA band structure  gives the  minimal model that should  be used 
to describe the physics of system under consideration.
Although LDA calculation does not give correct description of the SCES band structure, 
it gives {\it ab initio} parameters and reduced number of essential orbitals 
or the ``minimal reliable model''. Then,  the  effects  of  strong electron  
correlations in the framework of this model with {\it ab initio} calculated parameters are 
explicitly  taken  into  account within the GTB method and the quasiparticle band structure is derived.

In Section~\ref{sec_lda_par} the {\it ab initio} calculations 
for undoped La$_2$CuO$_4$ and Nd$_2$CuO$_4$ are presented.  
One can see that in the 3-band model (Figs.~\ref{la_x2_O} and \ref{nd_x2_O}) it is possible to describe 
the top of the valence band but not the lower lying excitations withing 4 eV.
The main effect of taking into account 
Cu-$d_{3z^2-r^2}$ and O$_a$-$p_z$ states for La$_2$CuO$_4$ 
system is the proper description of the band structure (in comparison with LDA calculation) 
at the energies up to 4 eV below Fermi level (see Fig.~\ref{la_x2_z2_O}).
Of course, the {\it ab initio} LDA band structure is not correct in undoped cuprates, but it 
gives an indication what orbitals should be included in more appropriate calculations. 
Therefore if one needs to describe quantitatively the low-energy excitations of 
La$_{2-x}$Sr$_x$CuO$_4$, the Cu-$d_{3z^2-r^2}$ and O$_a$-$p_z$ orbitals should 
be taken into account and the reliable minimal model is the multiband $p-d$ model.
In Nd$_2$CuO$_4$ the Cu-$d_{3z^2-r^2}$ and O2-$p_z$ states does not contribute 
significantly to the band structure (compare Figs.~\ref{nd_x2_O} and \ref{nd_x2_z2_O}) 
and the minimal model is the 3-band $p-d$ model. Nevertheless to treat p- and n-type 
cuprates on equal footing later we will use the same multiband $p-d$ model for both 
LSCO and NCCO with different material dependent parameters.
Hopping parameters decay rapidly with distance (see Tables~\ref{hopsLa} 
and ~\ref{hopsNd}) so in GTB calculation we will use only nearest copper-oxygen and 
oxygen-oxygen hoppings which are listed in Table~\ref{microscopicparams_table}.

In Refs.~\onlinecite{andersen1,pavarini1} {\it ab initio} calculations 
were done for YBa$_2$Cu$_3$O$_7$ and La$_2$CuO$_4$, and single-electron 
energy $\varepsilon_{p_x}=0.9$ eV was obtained. This value is very close 
to the one presented in Table~\ref{microscopicparams_table}. 
But in Refs.~\cite{andersen1,pavarini1} the Cu-$s$ states 
were taken into account with energy $\varepsilon_{s}=-6.5$ eV. 
Our LDA calculations shows that Cu-$s$ bands contributes to the 
band structure shown in Figs.~\ref{la_x2_z2_O} and \ref{nd_x2_z2_O} 
at approximately 7 eV below and at 2 eV above Fermi level. Therefore 
Cu-$s$ states does not contribute significantly to the low-energy physics. 
But these states can contribute to the effective intraplane hopping 
parameters $t'$ and $t''$ between the nearest and next-nearest neighboring 
unit cells. In our LDA+GTB method Cu-$s$ states are neglected. It could be a 
reason why for La$_2$CuO$_4$ our $t'/t=-0.137$ (see Table~\ref{ptypeparams_table}) 
is less then $t'/t=-0.17$ obtained in Ref.~\onlinecite{pavarini1}, where 
influence of Cu-$s$ orbital on hoppings was taken into account.

There is a claim that $pd\pi$-bonds \cite{moskvin1} and 
non-bonding oxygen states \cite{moskvin2} are very important 
in low-energy physics of High-$T_c$ cuprates. 
To discuss this topic lets start with analysis of {\it ab initio} calculations. 
Present LDA calculations show that anti-bonding bands $\pi$* of 
$\pi$-bonds (Cu-$d$+O-$p\pi$, see Figs.~\ref{la_x2_O} 
and \ref{nd_x2_O}) situated slightly below anti-bonding $\sigma_a$* 
bands of Cu-$d_{3z^2-r^2}$+O$_a$-$p_z$ origin in La$_2$CuO$_4$ and 
slightly above anti-bonding $\sigma_a$* bands of Cu-$d_{3z^2-r^2}$ 
origin in Nd$_2$CuO$_4$ (see Figs.~\ref{la_x2_z2_O} and \ref{nd_x2_z2_O}). 
GTB calculations \cite{gavrichkov1} show that states corresponding 
to $\sigma_a$* band contributes to the $a_{1g}$ molecular orbital in 
the single-hole sector of the Hilbert space. This $a_{1g}$ molecular orbital situated above 
$b_{1g}$ state $\left|\sigma\right> = \left\{ \left|\uparrow\right>, 
\left|\downarrow\right> \right\}$ by an energy about 1.2 eV. From the 
relative position of $\sigma_a$* and $\pi$* bands in LDA calculations 
we conclude that the energy of molecular orbital corresponding to the 
$\pi$* band will be situated around energy of $a_{1g}$ state. Therefore, 
it will be above $\left|\sigma\right>$ state by about $1.0 \div 1.4$ eV. 
Also, both states corresponding to $\pi$* and $\sigma_a$* are empty in 
undoped compound and spectral weight of quasiparticle excitations to or 
from these states will be zero. Summarizing, $\pi$-bonds, as well as 
$\sigma_a$* states, will contribute to the GTB dispersion only upon 
doping and only in the depth in the valence band below 1 eV from the top. 
Moreover, since energy difference between triplet $\left|T\right>$ and 
singlet $\left|S\right>$ states is about 0.5 eV, \cite{gavrichkov1} the 
contribution from the singlet-triplet excitations will be much more 
important to the low-energy physics. Although, $\pi$-bonds could be 
important for explanation of some optical and electron-energy loss 
spectroscopy experiments, but in description of low-energy physics of 
interest they could be neglected. The non-bonding oxygen states contribute 
to the valence band with energy about $2 \div 3$ eV below the top. That is 
why we will not take $\pi$-bonds and non-bonding oxygen states in our 
further consideration.

Now we have an idea what model should be used and {\it ab initio} microscopic parameters 
of this model.
As described in Section~\ref{sec_gtb}, the GTB method is appropriate method for 
description of SCES in Mott-Hubbard type insulators and it's results are in good agreement 
with experimental data. Then it is natural to use this method to work with 
the {\it ab initio} derived multiband $p-d$ model.

The parameters (\ref{eq:t}) of the Hamiltonian in the GTB
method derived from {\it ab initio} one are presented in 
Tables~\ref{ptypeparams_table} and \ref{ntypeparams_table} for p- and n-type
cuprates, respectively. 
Single-electron energies (in eV) and matrix elements of annihilation-creation 
operators in the X-operators representation were calculated for both LSCO:
\begin{eqnarray}
\label{eq:epsilonLSCO}
 &&\epsilon_1=-1.919, \epsilon_{2S}=-2.010, \epsilon_{2T}=-1.300,\nonumber\\
 &&u=-0.707, v=-0.708, \gamma_x=-0.619, \\
 &&\gamma_b=-0.987, \gamma_a=-0.032, \gamma_p=-0.962, \gamma_z=-0.237,\nonumber
\end{eqnarray}
and NCCO:
\begin{eqnarray}
\label{eq:epsilonNCCO}
 &&\epsilon_1=-1.660, \epsilon_{2S}=-1.225, \epsilon_{2T}=-0.264,\nonumber\\
 &&u=-0.756, v=-0.655, \gamma_x=0.626, \\
 &&\gamma_b=0.984, \gamma_a=-0.008, \gamma_p=0.997, \gamma_z=0.037.\nonumber
\end{eqnarray}

\begin{table}
\caption {\label{ptypeparams_table} Parameters of the multiband Hubbard model 
(\ref{eq:HX}) and exchange integral $J$ for LSCO obtained in the 
framework of the LDA+GTB method (all values in eV). Hoppings giving 
main contribution to the top of the valence band are shown by bold type.}
\begin{ruledtabular}
\begin{tabular}{c|rrrrrr}
$\rho$&
$t^{00}_{\rho}$&$\mathbf{t^{SS}_{\rho}}$&$t^{0S}_{\rho}$&$t^{TT}_{\rho}$&$t^{ST}_{\rho}$&$J_{\rho}$ \\
\hline
(0,1)  & 0.453 &\bf  0.679 & 0.560 & 0.004 &-0.086 & 0.157  \\
(1,1)  &-0.030 &\bf -0.093 &-0.055 &-0.001 & 0     & 0.001 \\
(0,2)  & 0.068 &\bf  0.112 & 0.087 & 0.002 &-0.016 & 0.004 \\
(2,1)  & 0.003 &\bf -0.005 & 0     & 0     &-0.002 & 0 \\
\end{tabular}
\end{ruledtabular}
\end{table}

\begin{table}
\caption {\label{ntypeparams_table} The same as in Table~\ref{ptypeparams_table}, but for NCCO. 
Hoppings giving main contribution to the bottom of the conductivity band are 
shown by bold type.}
\begin{ruledtabular}
\begin{tabular}{c|rrrrrr}
$\rho$& $\mathbf{ t^{00}_{\rho}}$&$t^{SS}_{\rho}$&$t^{0S}_{\rho}$&$t^{TT}_{\rho}$&$t^{ST}_{\rho}$&$J_{\rho}$ \\
\hline
(0,1)  &\bf  0.410 & 0.645 &-0.523 & 0 &-0.0052 & 0.137 \\
(1,1)  &\bf -0.013 &-0.076 & 0.035 & 0 & 0      & 0.001 \\
(0,2)  &\bf  0.058 & 0.104 &-0.078 & 0 &-0.0002 & 0.003 \\
(2,1)  &\bf  0.005 &-0.002 &-0.003 & 0 &-0.0004 & 0 \\
\end{tabular}
\end{ruledtabular}
\end{table}

It is known\cite{tohyama1} that sign of the hoppings in the $t-t'-t''-J$ model 
changes during electron-hole transformation of the operators. Therefore, $t_{\rho}$ 
will have different signs in p- and n-type cuprates. In present paper we don't do 
electron-hole transformation of the operators and both $t-t'-t''-J$* and singlet-triplet 
$t-t'-t''-J$* models are written using hole operators. Because of that there is no difference 
in signs of the hoppings $t_{\rho}$ for the hole and electron doped systems presented 
in Tables~\ref{ptypeparams_table} and \ref{ntypeparams_table}.

\begin{figure*}
\includegraphics[width=0.45\linewidth]{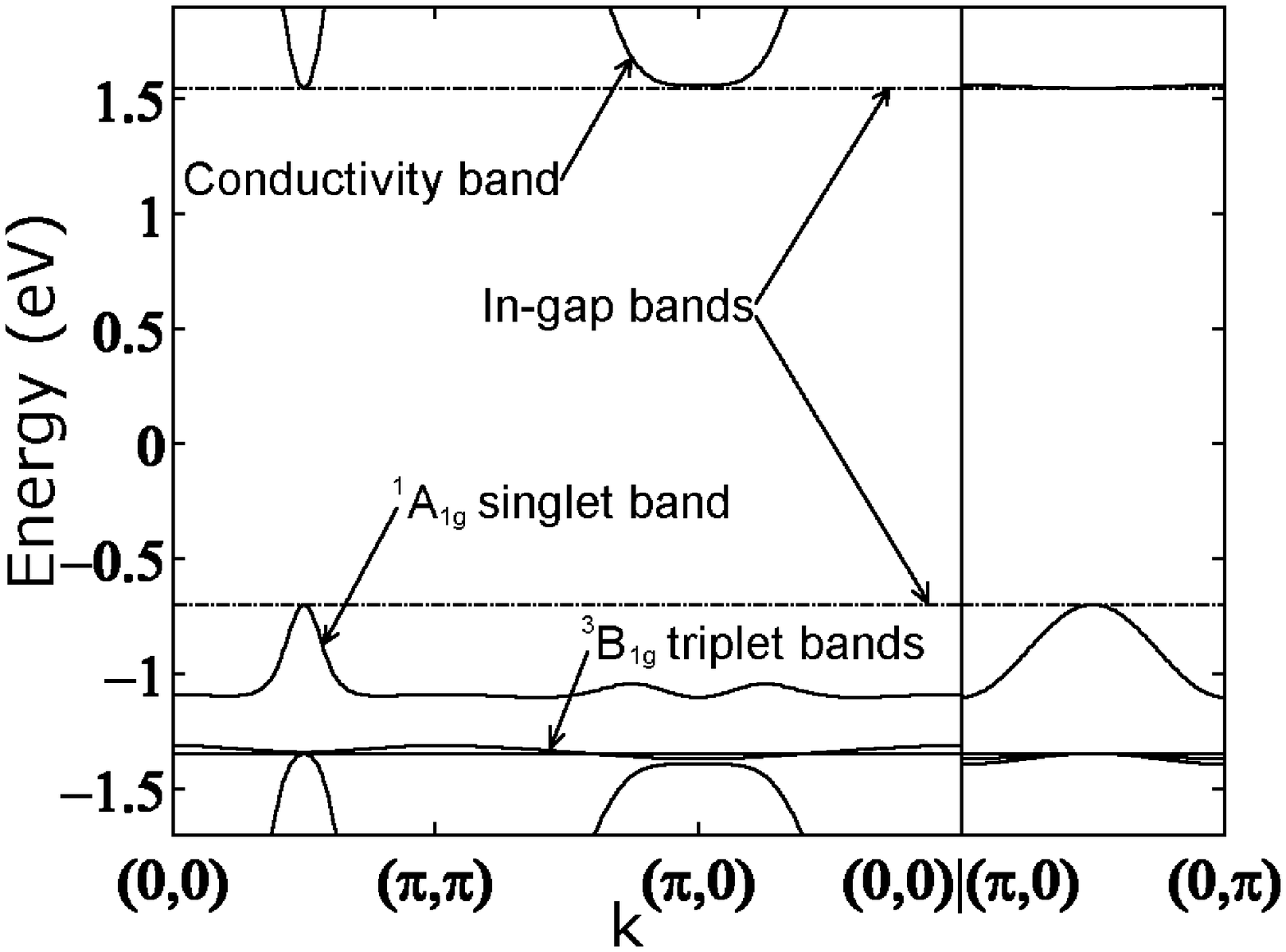}
\includegraphics[width=0.45\linewidth]{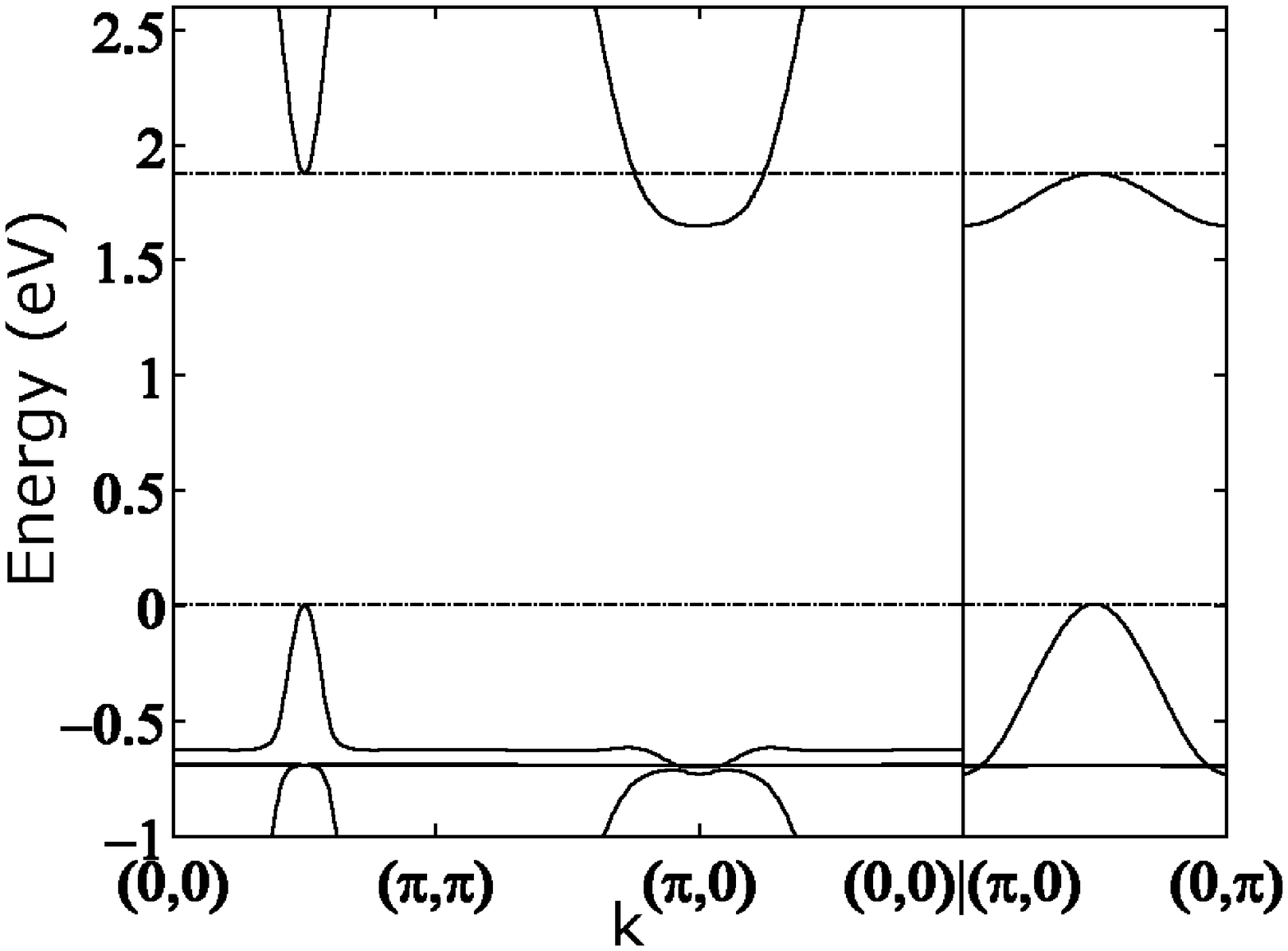}
\caption {\label{fig_compare_bands_p} 
The AFM band structure of La$_2$CuO$_4$ obtained in the GTB method with phenomenological 
set of parameters (left) and in the LDA+GTB method (right). 
In the left figure bands are labelled; in other GTB band structure  
figures relative positions of bands are the same.}
\end{figure*}

\begin{figure*}
\includegraphics[width=0.45\linewidth]{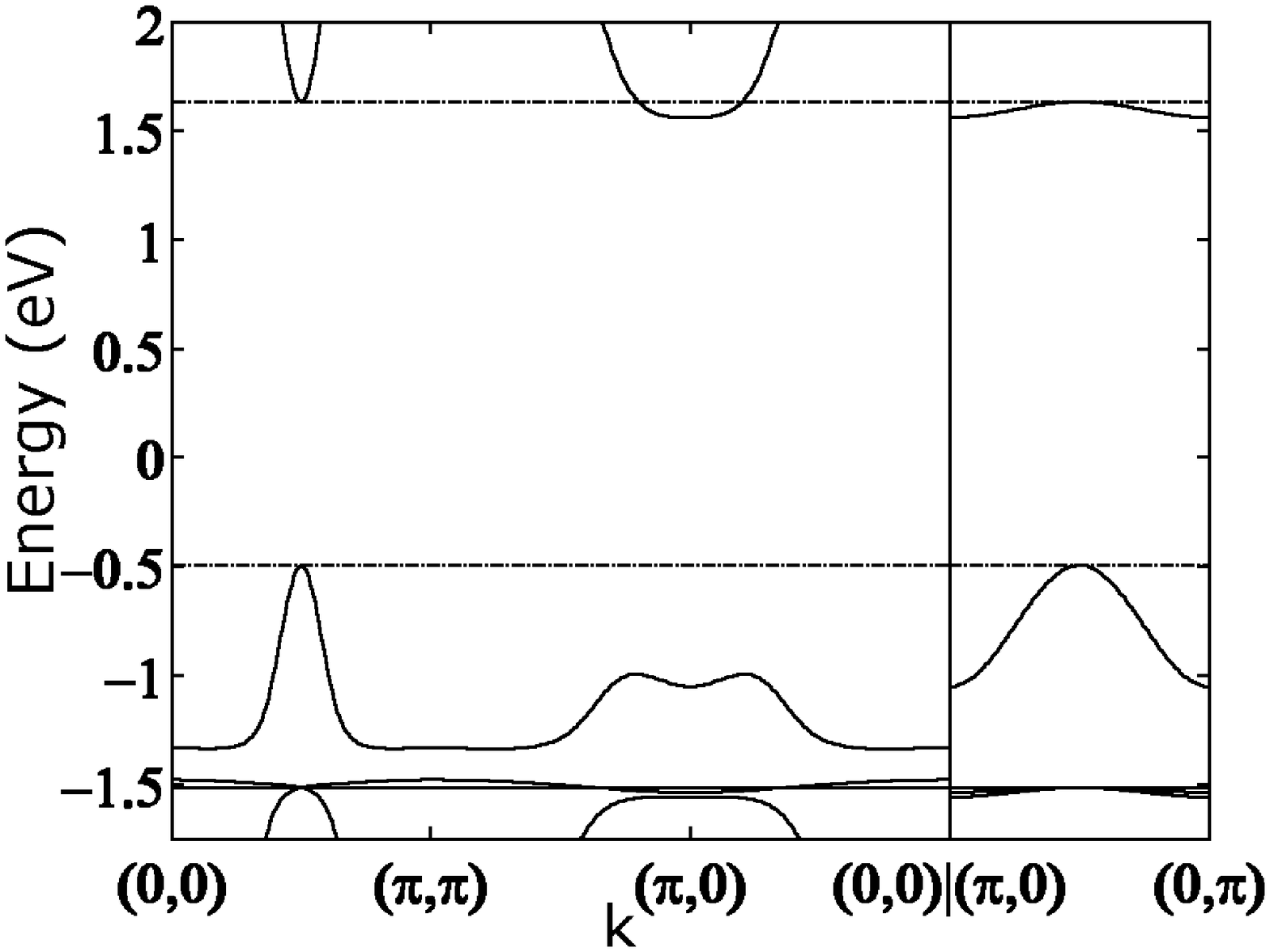}
\includegraphics[width=0.45\linewidth]{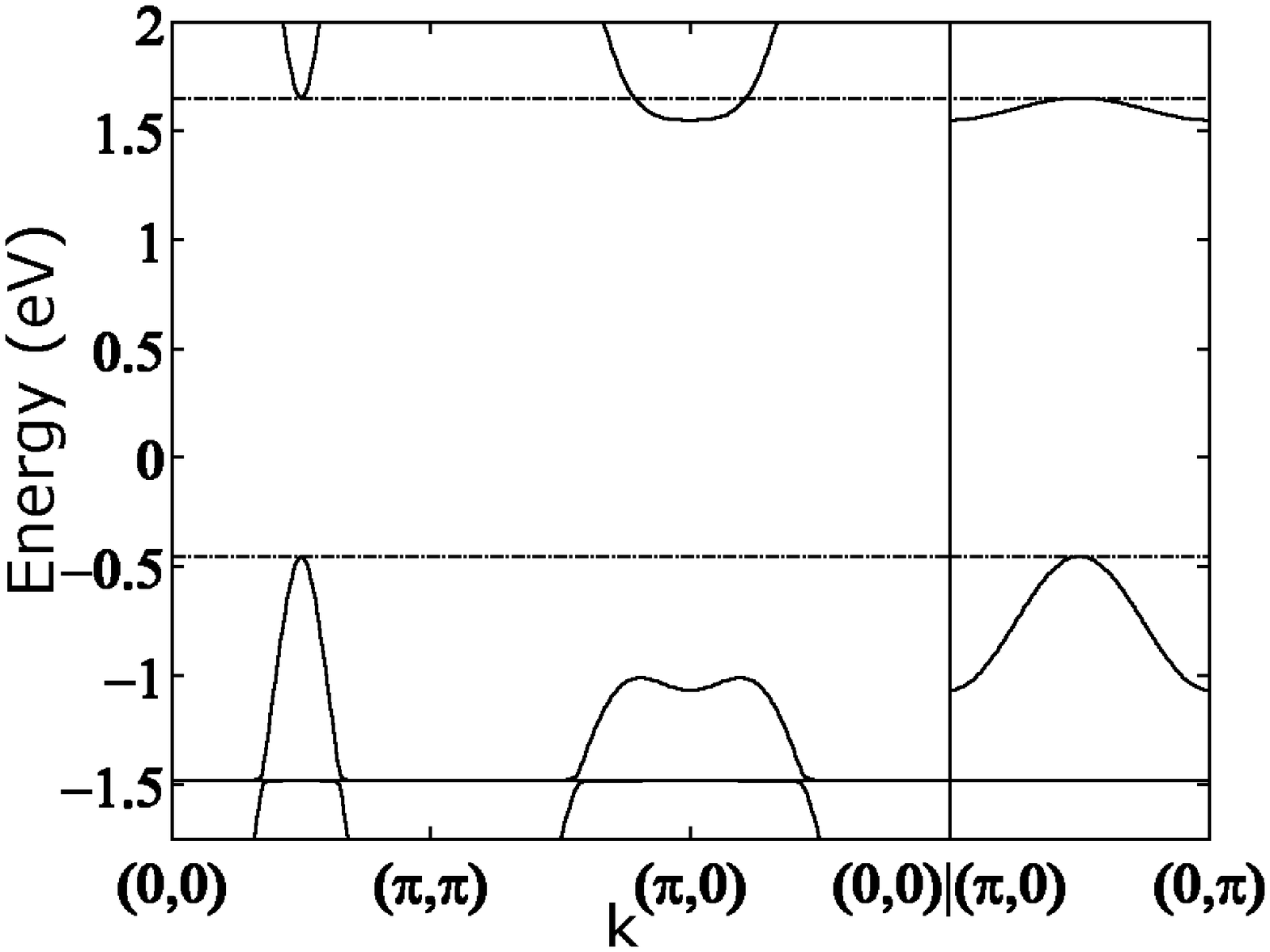}
\caption{\label{fig_compare_bands_n} 
The AFM band structure of Nd$_2$CuO$_4$ obtained in the GTB method with phenomenological 
set of parameters (left) and in the LDA+GTB method (right).}
\end{figure*}

As the next step we calculate the band structure of the undoped antiferromagnetic (AFM) 
insulating cuprate within the GTB method. 
Results for both GTB method with fitting parameters and LDA+GTB method 
with {\it ab initio} parameters (Table~\ref{microscopicparams_table}) are presented in the 
Fig.~\ref{fig_compare_bands_p} for La$_2$CuO$_4$ and in the Fig.~\ref{fig_compare_bands_n} 
for Nd$_2$CuO$_4$.

The GTB band structure obtained for both phenomenological and {\it ab initio} sets of 
parameters is almost identical: the valence band, located below 0 eV in figures, 
and the conductivity band, located above +1.5 eV, divided by the insulator gap of the 
charge transfer origin $E_{ct} \approx 2$ eV; 
the undoped La$_2$CuO$_4$ and Nd$_2$CuO$_4$ are insulators in both antiferromagnetic 
and paramagnetic states. In-gap states at the top of the valence band and about the 
bottom of the conductivity band are shown by dashed lines. Their spectral weights and 
dispersions are proportional to doping $x$ and concentration of magnons \cite{korshunov4}. 
Therefore, for undoped compounds, in the Hubbard-I approximation used in GTB method, 
these states are dispersionless with zero spectral weight.

The valence band have bandwidth about 6 eV and consists of a set of very narrow subbands 
with the highest one at the top of the valence band - the so-called ``Zhang-Rice singlet'' 
subband.
The dominant spectral weight in the singlet band stems 
from the oxygen $p$-states, while for the bottom of the empty conductivity band it is from 
$d_{x^2-y^2}$-states of copper.
Both methods give small, less then 0.5 eV, 
splitting between the $^1A_{1g}$ Zhang-Rice-type singlet band 
and $^3B_{1g}$ narrow triplet band located 
below the singlet band ({\it e.g.} in the Fig.~\ref{fig_compare_bands_n} 
for Nd$_2$CuO$_4$ it is located at -1.5 eV). 
The energy of Cu-$d_{3z^2-r^2}$ orbital plays the dominant role 
in this splitting in the GTB method. For La$_2$CuO$_4$ 
energy $\varepsilon_{d_{z^2}}$ is smaller then for Nd$_2$CuO$_4$ 
(see Table~\ref{microscopicparams_table}). This results in smaller 
width of the singlet band for the LSCO compared to the 
NCCO: about 0.5 eV and 1 eV correspondingly.

However, for La$_2$CuO$_4$ minor discrepancies occurs in the dispersion of the 
bottom of the conductivity band near $(\pi,0)$ point
obtained by GTB with phenomenological set of parameters and by LDA+GTB. 
This leads to the different character of the optical absorption edge in two presented methods. 
The absorption edge for the LDA+GTB is formed by the 
indirect transitions in contrast to the GTB method with phenomenological set of parameters, 
where the momentum of excited quasiparticle is conserved by optical transition at the 
absorption edge. For Nd$_2$CuO$_4$ both GTB method with fitting parameters and LDA+GTB
result in the conductivity band minima 
at the $(\pi,0)$ point (see Fig.~\ref{fig_compare_bands_n}).
Also, in the LDA+GTB method the triplet band dispersion and the singlet-triplet 
hybridization are much smaller then in the GTB method with fitting parameters. This 
happens mainly due to the smaller values of $t'_{pp}$ used in LDA+GTB method, because 
it is this microscopic parameter that gives main numerical contribution (see Eqs.~(\ref{eq:t}), 
(\ref{eq:epsilonLSCO}) and (\ref{eq:epsilonNCCO})) to the $t_{fg}^{TT}$ and 
$t_{fg}^{ST}$ - hoppings that determines the triplet band dispersion and the 
singlet-triplet hybridization respectively. 
So, despite some minor discrepancies, both GTB method with phenomenological 
parameters and LDA+GTB method without free parameters gives similar band dispersion.

Next topic that we will discuss in connection to the LDA+GTB method is the value of magnetic 
moment on copper $M_{\textrm{Cu}}$. From the neutron diffraction 
studies of La$_2$CuO$_4$ \cite{vaknin1} and YBa$_2$Cu$_3$O$_6$ \cite{tranquada1} it is 
known that $M_{\textrm{Cu}}$ is equal to $0.5 \mu_B$ where $\mu_B$ is Bohr magneton. There are 
two reasons of why $M_{\textrm {Cu}}$ is different from the free atomic value $1.14 \mu_B$ in $S=1/2$ Cu$^{2+}$,
namely zero temperature quantum spin fluctuations and the covalent effect. 
Since each oxygen have two neighboring coppers belonging to different magnetic sublattices the total 
moment on oxygen is equal to zero. But due to $p-d$ hybridization the $p$-states of 
oxygen are partially filled so these orbitals could carry non-zero magnetic moment 
$M_{\textrm {O}}$, while total moment on oxygen will be equal to zero. Such space distribution 
of magnetic moment leads to the difference \cite{freltoft1} between experimentally 
observed antiferromagnetic form-factor for La$_2$CuO$_4$ and the Heisenberg 
form-factor of Cu$^{2+}$. 
In order to take into account covalent effects and zero quantum fluctuations on equal 
footing we will write down the expression for $M_{\textrm {Cu}}$:
\begin{equation}
\label{eq:MCu}
M_{\textrm {Cu}} = 2.28 \mu_B \left< S^z \right> u^2,
\end{equation}
where zero quantum spin fluctuations are contained in $\left< S^z \right>$ and covalent 
effects are described by the weight $u^2$ of the $d^9p^6$ configuration. The last quantity 
is calculated in the framework of the LDA+GTB method and equal to $u^2=0.5$.
In paper~\onlinecite{horsch1} the value $\left< S^z \right> = 0.3$ was obtained self-consistently 
in the effective quasi-two-dimensional Heisenberg antiferromagnetic model for typical 
in $La_2 CuO_4 $ ratio $10^{-5}$ of the interplane and intraplane exchange parameters. 
Close value of $\left< S^z \right> = 0.319$ was obtained in Ref.~\onlinecite{katanin1} 
where also the plaquette ring-exchange was considered in Heisenberg Hamiltonian.
Using Eq.~(\ref{eq:MCu}) and above values of $u^2$ and $\left< S^z \right>$ we have 
calculated magnetic moment on copper $M_{\textrm {Cu}}=0.4 \mu_B$, that is close to the 
experimentally observed $M_{\textrm {Cu}}=0.5 \mu_B$.

Summarizing this section, we can conclude that the proposed LDA+GTB scheme works quite well and could 
be used for quantitative description of the High-$T_c$ cuprates. 
The LDA+GTB scheme also can be used for wide class of SCES - cuprates, manganites, and other.

\section{Effective low-energy model \label{sec_effective_model}}

When we are interested in the low-energy physics (like {\it e.g.} superconductivity) 
it is useful to reduce the microscopic model to more simpler effective Hamiltonian. 
For example, for the Hubbard model in the regime of strong correlations the effective 
model is the $t-J$* model ($t-J$ model plus 3-centers correlated hoppings $H_3$) 
obtained by exclusion of the intersubband 
hoppings perturbatively. \cite{bulaevskii1,chao1,hirsch1}
Analysis of the 3-band model results in the effective Hubbard and the $t-J$ 
model. \cite{rice1,lovtsov1,schutler1,belinicher1,feiner1}

As the next step we will formulate the effective model for the multiband $p-d$ model. 
Simplest way to do it is to neglect completely contribution of two-particle triplet state
$^3B_{1g}$. Then there will be only one low-energy two-particle state --
Zhang-Rice-type singlet $^1A_{1g}$ -- and the effective model will be the
usual $t-J$* model. But in the multiband $p-d$ model the difference
$\epsilon_T-\epsilon_S$ between energy of two-particle singlet and
two-particle triplet depends strongly on various model parameters,
particularly on distance of the apical oxygen from the planar oxygen,
energy of the apical oxygen, difference between energy of
$d_{z^2}$-orbitals and $d_{x^2}$-orbitals. For the realistic values of
model parameters $\varepsilon_T-\varepsilon_S$ is close to $0.5$~eV
\cite{gavrichkov2,raimondi1} contrary to the 3-band model with this value
being about $2$~eV. To take into account triplet states we will derive the 
effective Hamiltonian for multiband $p-d$ model by exclusion of the 
intersubband hopping between low (LHB) and upper (UHB) Hubbard subbands. These subbands 
divided by the energy of charge-transfer gap $E_{ct} \approx 2$ eV (similar to $U$ in the 
Hubbard model) and using perturbation theory, 
similar to Ref.~\onlinecite{chao1}, with small parameter $W/U$ we can derive 
separate effective models for UHB and LHB. This procedure is schematically shown 
in Fig.~\ref{fig_model_basis}. And, as one can see, since the UHB and LHB in 
initial model (\ref{eq:HXpd}) are formed by different quasiparticles 
(namely, $\alpha_0$ for LHB and $\alpha_1$, $\alpha_2$, $\alpha_3$ for UHB in
Fig.~\ref{fig_model_basis}), the effective models will be different for upper 
(valence band, hole doped) and lower (conductivity band, electron doped) subbands.

We write the Hamiltonian in the form $H=H_0+H_1$, where the excitations via
the charge transfer gap $E_{ct}$ are included in $H_1$. Then we define an
operator $H(\epsilon)=H_0+\epsilon H_1$ and make the unitary transformation
$\tilde{H}(\epsilon) = \exp{(-i \epsilon \hat{S})} H
\left(\epsilon\right) \exp{( i \epsilon \hat{S})}$. Vanishing
linear in $\epsilon$ component of $ \tilde{H}(\epsilon) $ gives the
equation for matrix $\hat{S}$: $H_1 + i \left[ H_0 ,\hat{S} \right]=0$. The
effective Hamiltonian is obtained in second order in $\epsilon$ and at
$\epsilon=1$ is given by $\tilde {H}=H_ 0+\frac {1}{2} i \left[ H_ 1, \hat {S} \right]$.
For the multiband $p-d$ model (\ref{eq:HXpd}) in case of electron doping we 
obtain the usual $t-J$* model describing conductivity band:
\begin{eqnarray}
\label{eq:HtJ}
H_{t-J*}&=&\sum_{f, \sigma} \varepsilon_1 X_{f}^{\sigma, \sigma} + 
\sum_{f \neq g, \sigma} t_{fg}^{0 0} X_{f}^{\sigma, 0} X_{g}^{0, \sigma}  + 
\sum\limits_{f \neq g \neq m, \sigma} H_3 \nonumber \\
&+& \sum_{f \neq g} J_{fg} \left( \vec{S}_f \vec{S}_g - \frac{1}{4} n_f n_g \right),
\end{eqnarray}
here $H_3$ contains three-centers interaction terms given 
by Eq.~(\ref{eq:H_3}), $\vec{S}_f$ are spin operators and $n_f$ are number of 
particles operators. The $J_{fg} = 2 \left( t_{fg}^{0S} \right)^{2} / E_{ct}$ is 
the exchange parameter.

For p-type systems the effective Hamiltonian has the form of the singlet-triplet $t-J$* model 
describing valence band:
\begin{eqnarray}
\label{eq:Heff}
H_{eff} &=& H_0 + H_t + \sum\limits_{ f \neq g \neq m, \sigma } H_{eff3} \nonumber \\
&+& \sum_{f \neq g}J_{fg} \left( \vec{S}_f \vec{S}_g - \frac{1}{4} n_f n_g \right).
\end{eqnarray}
Three-centers interaction terms $H_{eff3}$ are given by Eq.~(\ref{eq:H_eff3}). 
Expressions for $H_0$ and $H_t$ are as the follows:
\begin{eqnarray*}
H_0 &=& \sum_{f} \left[ \varepsilon_{1} \sum_{\sigma} X_{f}^{\sigma, \sigma}
+ \varepsilon_{2S} X_{f}^{S, S} + \varepsilon_{2T} \sum_{M} X_{f}^{TM, TM}
\right], \\
H_t &=& \sum_{f \neq g,\sigma} \Bigl\{ t_{fg}^{SS} X_{f}^{S,
\bar{\sigma}} X_{g}^{\bar{\sigma}, S}\\ &+& t_{fg}^{TT} \left( \sigma
\sqrt{2} X_{f}^{T0, \bar{\sigma}} - X_{f}^{T2\sigma, \sigma} \right) \left(
\sigma \sqrt{2} X_{g}^{\bar{\sigma}, T0} - X_{g}^{\sigma, T2\sigma} \right)\\
&+& t_{fg}^{ST} 2 \sigma \gamma_{b} \left[ X_{f}^{S, \bar{\sigma}} \left(
\sigma \sqrt{2} X_{g}^{\bar{\sigma}, T0} - X_{g}^{\sigma, T2\sigma} \right)
+h.c. \right] \Bigl\}.
\end{eqnarray*}

The resulting Hamiltonian (\ref{eq:Heff}) is the two-band generalization of the
$t-J$* model. Significant feature
of effective singlet-triplet model is the asymmetry for n- and p-type
systems which is known experimentally. So, we can conclude that for n-type
systems the usual $t-J$* model takes place while for p-type superconductors
with complicated structure on the top of the valence band the
singlet-triplet transitions are important.

\begin{figure}
\includegraphics[width=0.99\linewidth]{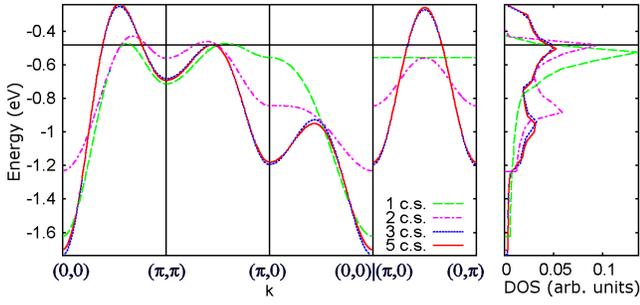}
\caption {\label{fig_tJ3model}
(Color online) Quasiparticle dispersion and corresponding densities of states (DOS) in 
the $t-J$* model calculated for different number of taken into account 
coordination spheres (c.s.). Chemical potential shown by the straight horizontal 
line was calculated self-consistently assuming 15\% holes doping.}
\end{figure}

Contrary to the multiband $p-d$ model's parameters that fall with distance rapidly, 
effective model parameters do not decrease so fast. This happens due to weak 
distance dependence of Wannier functions that determine coefficients $\mu_{fg}$, 
$\nu_{fg}$, $\lambda_{fg}$, $\xi_{fg}$, $\chi_{fg}$ \cite{gavrichkov1} which, 
in turn, determine distance dependence of effective model parameters (\ref{eq:t}).

To demonstrate the importance of hoppings to far coordination spheres (c.s.) in 
the Fig.~\ref{fig_tJ3model} we present the dispersion and DOS in the $t-J$* model 
with parameters from Table~\ref{ptypeparams_table}. The electron Green function 
(\ref{eq:D}) has been calculated beyond the Hubbard-I approximation by a decoupling 
of static correlation functions that includes short-range magnetic order: 
$\left<X_f^{\sigma \sigma } X_g^{\sigma '\sigma '}\right> \to n_p^2 + 
\frac{\sigma}{\sigma'} \frac{1}{2}C_{fg}$, 
$\left< X_f^{\sigma \bar \sigma } X_g^{\bar \sigma \sigma } \right> = C_{fg}$. Here $n_p$ 
is the occupation factors of the single-particle state, $C_{fg} = 2\left< S_f^z S_g^z \right>$ 
are static spin correlation functions which were self-consistently 
calculated from the spin Green's functions in the 2D $t-J$ model. \cite{sherman1} As 
one can easily see from Fig.~\ref{fig_tJ3model}, the dispersion with 
hoppings only to nearest 
neighbors (1 c.s.) and to next-nearest neighbors (2 c.s., the so called $t-t'-J$* model) 
is quantitatively different around $(0,0)$ point and {\it qualitatively} 
different around $(\pi,0)$ point from the dispersion with 3 c.s. ($t-t'-t''-J$* model) and 
more coordination spheres taken into account. 

Recent ARPES experiments\cite{zhou1} show that the Fermi velocity $v_F=E_F/k_F$
is nearly constant for wide range of p-type materials and doping independent within an 
experimental error of 20\%. We have calculated this quantity in the $t-J$* model with 
parameters from Table~\ref{ptypeparams_table} in the approximation described above. In 
the doping range from $x=0.03$ to $x=0.15$ our calculations give very weak doping dependence 
of the Fermi velocity. Assuming the lattice constant equal to 4{\AA} we have $v_F$ varying 
from $1.6$~eV{\AA}$^{-1}$ to $2.0$~eV{\AA}$^{-1}$. Taking into account experimental error 
of 20\% our results is very close to the experimental one.

\section{Conclusion \label{sec_conclusion}}

The approach developed here assumes the multiband 
Hamiltonian for the real crystal structure and its mapping onto low-energy model. 
Parameters of the effective model (\ref{eq:t}) are obtained directly from {\it ab initio} multiband 
model parameters. The sets of 
parameters for the effective models (\ref{eq:HtJ}) and (\ref{eq:Heff}) 
are presented in Tables~\ref{ptypeparams_table} and \ref{ntypeparams_table} for 
p- and n-type cuprates, correspondingly.

The effective low-energy model appears to be the $t-t'-t''-J$* model (\ref{eq:HtJ}) 
for Nd$_2$CuO$_4$ and the singlet-triplet $t-t'-t''-J$* model (\ref{eq:Heff}) for La$_2$CuO$_4$.
There is almost no difference in the band dispersion with addition of numerically 
small hoppings to 4-th, 5-th, etc. neighbors.

Summarizing, we have shown that the hybrid LDA+GTB method incorporate the {\it ab initio} 
calculated parameters of the multiband $p-d$ model and the adequate treatment 
of strong electron correlations.

\begin{acknowledgments}
The authors would like to thank A.V. Sherman for very helpful discussions. 
This work was supported by Joint Integration Program of Siberian and 
Ural Branches of Russian Academy of Sciences, 
RFBR grants 05-02-16301, 04-02-16096, 03-02-16124 and 05-02-17244, RFBR-GFEN  grant 03-02-39024, 
program of the Presidium of the Russian Academy of Sciences (RAS) ``Quantum macrophysics''. 
Z.P. and I.N. acknowledges support
from the Dynasty Foundation and International
Centre for Fundamental Physics in Moscow program for young
scientists 2005, Russian Science Support Foundation program for
best PhD students and postdocs of Russian Academy of Science 2005.
\end{acknowledgments}

\appendix*

\section{Expressions for 3-centers Correlated Hoppings in Effective Models}

In the $t-J$* model (\ref{eq:HtJ}) the 3-centers correlated hoppings are given by:
\begin{equation}
\label{eq:H_3}
H_3 = {\frac{{t_{fm}^{0S} t_{mg}^{0S} }}{{E_{ct} }}\left( X_f^{\sigma 0} X_m^{\bar \sigma \sigma } X_g^{0\bar \sigma } - X_f^{\sigma 0} X_m^{\bar \sigma \bar \sigma } X_g^{0\sigma } \right)}.
\end{equation}

The three-centers interaction terms $H_{eff3}$ in the effective Hamiltonian (\ref{eq:Heff}) are much more complicated then in the $t-J$* model due to additional triplet and singlet-triplet contributions:
\begin{equation}
\label{eq:H_eff3}
H_{eff3}  = {\frac{{t_{fm}^{0S} t_{mg}^{0S} }}{{E_{ct} }}H_3^{SS}  - v\frac{{t_{fm}^{0S} t_{mg}^{ST} }}{{E_{ct} }}H_3^{ST}  + v^2 \frac{{t_{fm}^{ST} t_{mg}^{ST} }}{{E_{ct} }}H_3^{TT}},
\end{equation}

\begin{equation*}
H_3^{SS}  = \left( {X_f^{\sigma S} X_m^{\bar \sigma \sigma } X_g^{S\bar \sigma }  - X_f^{\bar \sigma S} X_m^{\sigma \sigma } X_g^{S\bar \sigma } } \right),
\end{equation*}

\begin{eqnarray*}
H_3^{ST} &=& \frac{1}{{\sqrt 2 }}\left( {X_f^{\sigma T0} X_m^{\bar \sigma \sigma } X_g^{S\bar \sigma } - X_f^{\bar \sigma T0} X_m^{\sigma \sigma } X_g^{S\bar \sigma } } \right) \\ 
&+& 2\sigma \left( {X_f^{\bar \sigma T2\bar \sigma } X_m^{\bar \sigma \sigma } X_g^{S\bar \sigma } + X_f^{\sigma T2\sigma } X_m^{\sigma \sigma } X_g^{S\bar \sigma } } \right) \\ 
&+& \frac{1}{{\sqrt 2 }}\left( {X_f^{\sigma S} X_m^{\bar \sigma \sigma } X_g^{T0\bar \sigma }  - X_f^{\bar \sigma S} X_m^{\sigma \sigma } X_g^{T0\bar \sigma } } \right) \\ 
&-& 2\sigma \left( {X_f^{\sigma S} X_m^{\bar \sigma \sigma } X_g^{T2\sigma \sigma }  - X_f^{\bar \sigma S} X_m^{\sigma \sigma } X_g^{T2\sigma \sigma } } \right),
\end{eqnarray*}

\begin{eqnarray*}
H_3^{TT} &=& \frac{1}{2}\left( {X_f^{\sigma T0} X_m^{\bar \sigma \sigma } X_g^{T0\bar \sigma }  - X_f^{\bar \sigma T0} X_m^{\sigma \sigma } X_g^{T0\bar \sigma } } \right) \\ 
&-& \left( {X_f^{\bar \sigma T2\bar \sigma } X_m^{\bar \sigma \sigma } X_g^{T2\sigma \sigma }  + X_f^{\sigma T2\sigma } X_m^{\sigma \sigma } X_g^{T2\sigma \sigma } } \right) \\ 
&+& \frac{{2\sigma }}{{\sqrt 2 }}\left( { - X_f^{\sigma T0} X_m^{\bar \sigma \sigma } X_g^{T2\sigma \sigma }  + X_f^{\bar \sigma T0} X_m^{\sigma \sigma } X_g^{T2\sigma \sigma } } \right) \\ 
&+& \frac{{2\sigma }}{{\sqrt 2 }}\left( { + X_f^{\bar \sigma T2\bar \sigma } X_m^{\bar \sigma \sigma } X_g^{T0\bar \sigma }  + X_f^{\sigma T2\sigma } X_m^{\sigma \sigma } X_g^{T0\bar \sigma } } \right).
\end{eqnarray*}

\end{document}